\input phyzzx
\def\e{\adveq\eqno{\rm (\chapterlabel\the\equanumber)}}
\def\mysec#1{\equanumber=0\chapter{#1}}
\def\adveq{\global\advance\equanumber by 1}
\def\myeq{{\rm \chapterlabel.\the\equanumber}}
\def\rarrow{\rightarrow}

\def\twoline#1#2{\displaylines{\qquad#1\hfill(\adveq\myeq)\cr\hfill#2
\qquad\cr}}
\def\enumber{\llap(\adveq\myeq)}

\def\manyeq#1{\eqalign{#1}\e}

\def\semidirect{\mathrel{\raise0.04cm\hbox{${\scriptscriptstyle |\!}$
\hskip-0.175cm}\times}}


\def\ref#1{$^{[#1]}$}

\def\pr#1{#1^\prime}
 
\def\r#1{$[\rm#1]$}

\def\threeline#1#2#3{\displaylines{\qquad#1\hfill\cr\hfill#2\hfill\llap{(\adveq\myeq)}\cr
\hfill#3\qquad\cr}}

\def\e{\adveq\eqno{\rm (\chapterlabel.\the\equanumber)}}
\def\mysec#1{\equanumber=0\chapter{#1}}
\def\adveq{\global\advance\equanumber by 1}
\def\myeq{{\rm \chapterlabel.\the\equanumber}}
\def\rarrow{\rightarrow}

\def\twoline#1#2{\displaylines{\qquad#1\hfill(\adveq\myeq)\cr\hfill#2
\qquad\cr}}
\def\enumber{\llap(\adveq\myeq)}

\def\manyeq#1{\eqalign{#1}\e}

\def\semidirect{\mathrel{\raise0.04cm\hbox{${\scriptscriptstyle |\!}$
\hskip-0.175cm}\times}}


\def\ref#1{$^{[#1]}$}

\def\pr#1{#1^\prime}
 
\def\r#1{$[\rm#1]$}

\def\threeline#1#2#3{\displaylines{\qquad#1\hfill\cr\hfill#2\hfill\llap{(\adveq\myeq)}\cr
\hfill#3\qquad\cr}}

\def\threeline#1#2#3{\displaylines{\qquad#1\hfill\llap{(\adveq\myeq)}\cr\hfill#2\hfill \cr
\hfill#3\qquad\cr}}

\input tikz.tex
\def\center#1\endcenter{\centerline{#1}}
\overfullrule=0pt

\def\frac#1#2{{#1\over #2}}
\def\approx{\sim}
\def\pr{\prime}

\font\mbb=msbm10 
\newfam\bbb
\textfont\bbb=\mbb

\date{August,  2019}
\date{August, 2019}
\titlepage
\title{The 4--CB Algebra and Solvable Lattice Models}
\author{Vladimir  Belavin$^{a,b,c}$, Doron Gepner$^f$, Jian-Rong Li$^d$ and  Ran Tessler$^e$}
\vskip20pt
\line{\it\hfill  $^a$\  Physics Department, Ariel University, Ariel 40700, Israel \hfill}
\line{\it \hfill $^b$ \  I.E. Tamm  Department of Theoretical Physics, P.N. Lebedev Physical
\hfill }\line{\hfill\it  Institute, Leninsky av. 53, 11991 Moscow, Russia\hfill}
\line{\it \hfill $^c$ \   Department of Quantum Physics, Institute for Information Transmission\hfill}
\line{\hfill\it  Problems, Bolshoy Karetny per. 19,  127994 Moscow, Russia\hfill}
\line{\hfill\it $^d$ \ Institute of Mathematics and Scientific Computing, University of Graz\hfill}
\line{\hfill\it Graz 8010, Austria\hfill}
\line{\hfill\it $^e$ \ Incumbent of the Lilian and George Lyttle Career Development Chair,\hfill}
\line{\hfill\it Department of Mathematics, Weizmann Institute,\hfill}
\line{\it\hfill Rehovot 76100, Israel\hfill}
\line{\it\hfill $^f$ Department of Particle Physics and Astrophysics, Weizmann Institute,\hfill}
 \line{\it\hfill Rehovot 76100,  Israel\hfill} 
 
 \endpage

\abstract

We study the algebras underlying solvable lattice models of the type fusion interaction round the
face (IRF). We propose that the algebras are universal, depending only on the number of blocks,
which is the degree of polynomial equation obeyed by the Boltzmann weights. Using the Yang--Baxter 
equation and the ansatz for the Baxterization of the models, we show that the three blocks models
obey a version of Birman--Murakami--Wenzl (BMW) algebra. For four blocks, we conjecture that the algebra is
the BMW  algebra with a different skein relation, along with one additional relation, and we provide evidence for this conjecture. We connect these 
algebras to knot theory by conjecturing new link invariants. 
The link invariants, in the case of four blocks, depend on three arbitrary parameters.
We check our result  for $G_2$ model
with the seven dimensional representation and for $SU(2)$  with the isospin $3/2$ representation, which are both four blocks theories.

\endpage 

\mysec{Introduction}

Solvable lattice models are important as exact models of statistical physics, conformal field
theory and phase transitions. For a review see \REF\Baxter{R.J. Baxter, Exactly solved models in statistical mechanics, Academic Press, London, U.K. (1982).} \r\Baxter. These models are also connected to
domains of  mathematics, being of importance in subjects such as Rogers Ramanujan identities,
knot theory and algebra.

Of particular interest, both in mathematics and physics, is the underlying algebraic structure of 
solvable lattice
models. 
Our general idea is that the algebra depends only on the number of blocks, which is 
the degree of the polynomial equation obeyed by the Boltzmann weights, and is general to all
the models, independently of the details of the model. This paper is directed towards proving this 
pivotal assertion.

We started  investigating this structure in the papers
\REF\First{V. Belavin and D. Gepner,  Nucl.Phys. B938 (2019) 223.}
\REF\Second{V. Belavin and D. Gepner, JHEP 1902 (2019) 033.}
\r{\First,\Second}, based on the initial results of the work
\REF\Found{D. Gepner, CALT-68-1825, hep-th/9211100 (1992).}
\r\Found. 
For two blocks, it is well known to be described by 
Templerley--Lieb algebra 
\REF\TL{ N. Temperley and E. Lieb, Proc.R.Soc. A 322 (1971) 251.}
\r\TL, or equivalently Hecke algebra.
We study here the three and four blocks cases. We base our investigation on an ansatz for the Baxterisation
put forward in ref. \r\Found. Using this ansatz and the Yang--Baxter equation (YBE), we show that
the three blocks case obeys a weak version of the Birman--Murakami--Wenzl algebra (BMW)
\REF\BW{J.S. Birman and H. Wenzl, Trans.Amer.Math.Soc. 313 (1989) 249.}
\REF\Mur{J. Murakami, Osaka.J.Math. 24 (1987) 745.}
\r{\BW,\Mur}. This is discussed in Section (2).

For the four blocks case, we find using the ansatz, an algebra which we call 4--CB (Conformal Braiding).
This algebra includes the BMW algebra, with a different skein relation, along with one
additional relation. This is described in Section (3). 
The connection of this algebraic structure to knot theory is 
described in Section (4).

We check the ansatz for $G_2$ theory, which is a four blocks theory, in Section (5). We already checked it for $SU(2)$ with the field of isposin $3/2$ in ref. \r\Second. We find that it holds in both cases.
We connect the algebraic structure with the tangle algebra of 
Kuperberg and Kalfagianni
\REF\Kuper{G. Kuperberg, The 1,0,1,1,4,10 Ansatz, Berkley preprint (1991).}
\REF\Kal{E. Kalfagianni, J. of Knot Theory and Its Ramifications, Vol. 2 No. 4 (1993) 431.}
\r{\Kuper,\Kal}. We find that Kalfagianni's relations hold for any four blocks theory,
assuming the ansatz, YBE and BMW. This is treated in Sections (6--7).

{\bf 1.1. Solvable IRF models.}

1.1.1. {\it Interaction-round-the-face (IRF) models.}

Let $I$ be a set, which is assumed to be finite for the purposes of this article.
We assume that $I$ is endowed with two relations $\approx_h,\approx_v.$
When $a\approx_h b$ ($a\approx_v b$) we say that $(a,b)$ are {\it horizontally (vertically) admissible}.
In the special case $\approx_h\equiv\approx_v,$ which will be mainly considered here, we write $\approx$ for $\approx_h,$$\approx_v$ and we say that $a,b$ are {\it admissible} whenever $a\approx b.$
The third piece of data we require is a function, called the {\it Boltzmann weight}, of four elements of $a,b,c,d\in I$ and a complex parameter $u,$ which is required to satisfy
$$\omega\left ( \matrix{a& b\cr c&d\cr}\bigg | u\right)=0,$$
unless the {\it admissibility condition} $$a\approx_h b,a\approx_v c,c\approx_h d,b\approx_v d\e$$ is met. The parameter $u$ is called the {\it spectral parameter}.

An interaction-round-the-face (IRF) lattice model is defined on a two dimensional square lattice, or its finite approximation via a $M\times M$ box with periodic boundary conditions (we will not make use in the periodicity in the algebraic treatment below, and will keep it only for the combinatorial motivation). We denote the underlying graph in both cases by $T$. A {\it configuration} is an assignment of an element of $I$ to each vertex of $T.$
The {\it partition function} of the model is defined to be
$$Z=Z(u)=Z_T(u):=\sum_{\rm configurations}\prod_{\rm faces} \omega\left ( \matrix{a& b\cr c&d\cr}\bigg | u\right),
\e$$
The {\it state space} of the theory is $({\bf C}^I)^{\otimes N},$
 and we denote states using the ket notation $\left|a_1\ldots a_N\right>.$ Dual states are denoted using the bra notation, $\left<a_1\ldots a_N\right|.$ A state is {\it admissible} if for all $i,~a_i\approx a_{i\pm 1}$ (we assume here $\approx_h=\approx_v,$ otherwise there are analogous requirements, but which depend on the parity of $i$).
Denote by $V_{adm}$ the space spanned by the admissible states.

1.1.2. {\it Solvability, Yang-Baxter equation (YBE) and braiding.}

We define the $i^{th}$ {\it face transfer matrix} $X_i(u)$ by
$$\left<a_1,a_2,\ldots,a_N | X_i(u) | a_1^\pr,a_2^\pr,\ldots,a_N^\pr\right>
=\left[\prod_{j\neq i}\delta_{a_j,a_j^\pr }\right] \omega\left(\matrix{a_{i-1} & a_i \cr a_i^\pr & a_{i+1} } \bigg | u\right).\e$$
For any $u,v,$ we have
$$X_i(u)X_{j}(v)=X_j(v)X_i(u),~~j\neq i\pm1.\e$$
The {\it Yang-Baxter equation (YBE)} is the relation
$$X_i(u) X_{i+1}(u+v) X_i(v)=X_{i+1}(v) X_i(u+v) X_{i+1}(u).\e
$$
An equivalent formulation of this equation, in terms of the Boltzmann weights, is
$$\twoline{\sum_c\omega\left(\matrix{g & c \cr a & b } \bigg | u\right)
\omega\left(\matrix{c & e \cr b & d } \bigg | u+v\right)
\omega\left(\matrix{g & f \cr c & e } \bigg | v\right)=}
{~~~~~~~~~~~~~=
\sum_c\omega\left(\matrix{a & c \cr b & d } \bigg | v\right)
\omega\left(\matrix{g & f \cr a & c } \bigg | u+v\right)
\omega\left(\matrix{f & e \cr c & d } \bigg | u\right).
}$$
If the YBE is satisfied then transfer matrices for different spectral parameters commute.

{\bf Remark 1.}
Although the point of view presented in this paper is algebraic, we make a small digression concerning the combinatorial picture. A {\it front} in a $M\times M$ box $T$ with periodic boundary conditions (or equivalently a $M\times M$ torus) is a chain of $N=2M-2$ vertices $v_1,\ldots,v_N$  such that $v_i$ is a neighbor of $v_{i+1},$ where addition is modulo $N,$ and such that the projections of $v_1,\ldots,v_N$ on the diagonal $x=-y$ are increasing, when the diagonal is oriented from NW to SE. An example of a front is the {\it standard front}, which is an arbitrary shift of the vertices $$(0,M),(0,M-1)(1,M-1),(1,M-2),\ldots(M,0)=(0,M)$$ by a lattice vector. A state can be thought of as an association of an element of $I$ to each vertex of the front. The state is admissible if neighboring elements are. Since all fronts are of the same size, there are isomorphisms between their state spaces. 

The $i^{th}$ face transfer matrix should be thought as promoting the front from $(v_1,\ldots,v_{i-1},v_i,v_{i+1},\ldots,v_N)$ to $(v_1,\ldots,v_{i-1},v_i,v_{i+1},\ldots,v_N)$ by adding a square whose four vertices are $$v_i,~~v_{i-1}=v_i+(1,0),~~v_{i+1}=v_i+(0,1),~~v_i,~~v'_i=v_i+(1,1).$$
We see that the face transfer matrices are operators between different, although isomorphic, state spaces.

Starting from a standard front, applying the {\it transfer matrix} $$X=X(u)=X_1(u)\cdot X_3(u)\cdot X_5(u)\cdots$$ amounts to pushing the front by $(1,0)$ and {\it cyclically shifting indices by} $1.$ Thus, roughly speaking $$Z_T(u)=Tr(X^{M}).$$
This means that if $\mu_1(u),\ldots,\mu_r(u)$ are the eigenvalues of $X,$ counted with multiplicities, the partition function equals
$$Z(u)=\sum_{i=1}^r \mu_i(u)^{M}.$$

If one can understand the eigenvalues of the transfer matrix good enough, the model is usually solvable, meaning that its partition function can be calculated. 
It follows from direct calculation that when the Yang-Baxter equation (1.5) and the commutation equation (1.4)  hold, the matrices $X(u)$ for different $u$ commute. This means that they have common eigenspaces. In many occasions the combinatorics or physics of the IRF model give rise to additional constrains on the transfer matrices, which in turn give rise to functional equations satisfies by the different $ \mu_i(u).$ Sometimes these constraints are strong enough to determine the eigenvalues.

One such constraint may be an inversion relation which connects $X_i(u)$ and $X_i(-u).$ We will consider such an inversion relation below. More details about the transfer matrix method in statistical mechanics models, as well as different inversion relations can be found in ref. \r\Baxter.

Suppose that the UV limit of the face transfer matrices
$$X_i=\lim_{u\rarrow i\infty} g(u) X_i(u),\e$$
exists and is finite and non zero, where $g(u)$ is some function.
Then the matrix coefficients of $X_i$ automatically satisfy the admissibility conditions equation (1.1). 
In addition, one can also take the limits of equations (1.4),(1.5) 
to obtain
$$X_iX_{j}=X_jX_i,~~j\neq i\pm1.\e$$
$$X_i X_{i+1} X_i=X_{i+1} X_i X_{i+1}.\e$$
These equations imply that $X_1,\ldots,X_N$ form a representation of the braid group. 

{\bf 1.2. RCFTs and Fusion IRF models}

In ref. \r\Found\ a conjectural recipe for obtaining solvable IRF models from a rational conformal field theory was described. We review it here, but refer to ref. \r\Found\ for more details.
We begin by providing a very partial definition of conformal field theories, and recalling the most basic properties we need for the discussion below. We refer the reader to ref. \REF\MS{G.W. Moore and N. Seiberg,  Phys. Lett. B 212 (1988) 451.}\r\MS\  for a complete definition and extensive analysis of conformal field theories.

A {\it conformal field theory (CFT)} $\cal O$ is a collection of {\it primary fields}, labelled by a set $I,$ together with a {\it fusion product}, which is a commutative and associative product defined via the fusion structure constants $f_{ab}^c$ ($a,b,c$ are elements of $I$)
$$a\times b=f_{ab}^c c.$$
Each primary field has a {\it conformal dimension} which is a non negative rational number which specifies the behaviour of the field under conformal symmetries.
We identify elements of $I$ with the corresponding primary fields. $\cal O$ is a {\it rational conformal field theory (RCFT)} if $I$ is finite.

Given a RCFT $\cal O$ and two fields $h,v$ one can use the fusion product to write admissibility conditions $\approx_h,\approx_v$ as follows:
$$a\approx_h b\Leftrightarrow  f_{ah}^b>0,~~a\approx_h b\Leftrightarrow  f_{av}^b>0.$$
An IRF model with a set of states $I$ and the admissibility conditions above is called a {\it fusion IRF} model.
Such a model is completely specified by its Boltzmann weights $\omega\left ( \matrix{a& b\cr c&d\cr}\bigg | u\right)$
which vanish unless
$$f_{a h}^b>0,~~f_{c h}^d > 0, ~~f_{a v}^c>0,~~\hbox{and}~~f_{b v}^d>0.\e$$

A fusion IRF model is called a {\it $n$ conformal braiding (CB) IRF} if
the fusion product of the primary fields $h$ and $v$ is a sum of $n$ primary fields
$$[h]\times [v]=\sum_{i=0}^{n-1} \psi_i.\e$$

1.2.1 {\it Braiding.}

We are interested in constructing solvable fusion IRF models, meaning models for which the Bolzmann weights satisfy (1.6). 
As we saw above, for such a model, if one can define the UV limit, a representation of the braid group appears, and it is a representation in which the matrix components satisfy the admissibility conditions (1.10).

To a conformal field theory there are associated {\it braiding matrices}
$$C_{c,d}\left[
\matrix{h & v\cr a &b\cr }\right].\e$$
The matrix components of these matrices vanish unless the admissibility conditions (1.10) 
 holds.
In addition, these matrices satisfy the hexagon relation, and when $h=v,$ this relation reduces to the braiding relations (1.8),(1.9).

Moreover, it can be shown that the matrix $C$ whose components are given by (1.12)  
for fixed $h,v$ satisfies the characteristic equation
$$\prod_{i=0}^{n-1} (C-\lambda_i)=0,$$ where $n$ is the number of blocks, and $\lambda_i$ are given by
$$\lambda_i=\epsilon_i e^{i\pi (\Delta_h+\Delta_v-\Delta_i)},\e$$
here $\Delta_i,\Delta_h,\Delta_v$ are the conformal dimensions of $\psi_i,h,v$ respectively, and $\epsilon_i=\pm 1$ according
to whether the product is symmetric or anti--symmetric.

1.2.2 {\it An ansatz for Baxterization.}

From now on we consider $h=\bar{h}=v.$
The appearance of the natural braiding matrix suggests searching for a solvable fusion IRF model with this matrix as the UV limit.
More precisely, we start with matrices $X_i,~i=1,\ldots, N,$ which are given by equation (1.3) with $\omega\left ( \matrix{a& b\cr c&d\cr}\bigg | u\right)$ replaced by
$C\left ( \matrix{a& b\cr c&d\cr}\right):=C_{c,d}\left[
\matrix{h & h\cr a &b\cr }\right]$.

These matrices satisfy the braiding equations (1.8),(1.9), 
the admissibility condition (1.10), 
and correspond to $n$ conformal blocks, meaning
$$\prod_{p=0}^{n-1} (X_i-\lambda_p)=0,
\e$$
where $\lambda_i$ are given in equation (1.13). 

The goal is to construct matrices $X_i(u),$ which satisfy the admissibility relation (1.10), equations (1.4),(1.5), 
 and that their UV limit is
$$\lim_{u\rarrow i\infty} g(u) X_i(u)=X_i
\e$$ for some function $g(u)$.
The process of extending a representation $X_i,~i=1,\ldots, N$ of the braid group, to a solution $X_i(u),$ of the Yang-Baxter equation (1.5) 
is called a {\it Baxterization}.

The ansatz of ref. \r\Found\ is the following.
Observe that
$$X_i=\sum_{a=0}^{n-1} \lambda_a P_i^a,
\e$$ where the projection $P_i^a$ to the $a^{th}$ eigenspace is given by
$$P_i^a=\prod_{p\neq a} \left[ {X_i-\lambda_p\over \lambda_a-\lambda_p}\right].
\e$$
These projections satisfy the relations
$$\sum_{a=0}^{n-1} P_i^a=1_i,\qquad P_i^a P_i^b=\delta_{a,b} P_i^b.
\e$$
{\bf Remark 2.}
The operator $1_i$ is just the identity on the space of admissible states, and $0$ on the space spanned by the complementary states.
In the algebraic analysis we will conduct in the following sections, since we will restrict only to admissible states, and the face transfer matrices preserve $V_{adm},$ we will be able to identify $1_i$ with the identity operator. 
The reason we still use the notation $1_i$ is that, as was explained in Remark 1 above, although we identify state spaces of different fronts with $V_{adm},$ combinatorially it is more accurate to consider our operators are relating state spaces of different fronts. With this point of view $1_i$ has the meaning of a unit face transfer matrix, which promotes the front from $(v_1,\ldots,v_{i-1},v_i,v_{i+1},\ldots,v_N)$ to $(v_1,\ldots,v_{i-1},v'_i,v_{i+1},\ldots,v_N)$ by adding a square whose four vertices are $$v_i,~~v_{i-1}=v_i+(1,0),~~v_{i+1}=v_i+(0,1),~~v_i,~~v'_i=v_i+(1,1),$$ and the field which is assigned to $v'_i$ in the admissible case is the same field that is assigned to $v_i.$

We define the {\it crossing parameters} as
$$\zeta_i=\pi (\Delta_{i+1}-\Delta_i)/2,~~\lambda=\zeta_0.\e$$

The trigonometric ansatz for the Yang Baxter equations (1.5) 
is
$$X_i(u)=\sum_{a=0}^{n-1} f_a(u) P_i^a,\e$$
where the functions $f_a(u)$ are given by
$$f_a(u)=\left[ \prod_{r=1}^a \sin(\zeta_{r-1}-u) \right]
               \left[ \prod_{r=a+1}^{n-1} \sin(\zeta_{r-1}+u) \right]\bigg/
               \left[ \prod_{r=1}^{n-1} \sin(\zeta_{r-1})\right] .\e$$
With this ansatz the following {\it inversion relation} or {\it unitarity} is straight forward
$$X_i(u) X_i(-u)=\rho(u) \rho(-u)1_i,
\e$$
where the function $\rho$ is defined by
$$\rho(u)=\prod_{r=1}^{n-1} {\sin(\zeta_{r-1}-u) \over \sin(\zeta_{r-1})}.
\e$$

It is also conjectured in ref. \r\Found\ that the Boltzmann weights obey crossing symmetry
$$\omega\left( \matrix {a & b\cr c& d\cr} \bigg | \lambda-u\right)=
\left[ {G_b G_c\over G_a G_d}\right]^{1/2} \omega \left( \matrix{ c & a \cr d &b \cr} \bigg| u\right),\e$$
where $G_a$ is some factor and $\lambda=\zeta_0$ is the crossing parameter.

The IRF model given by this ansatz is called the $({\cal O} ,h,v)$ fusion IRF model.

{\bf 1.3. $n$ Conformal Braiding (CB) Algebras.}

The algebras formed by the operators $X_i(u),~i=1,\ldots,N,$ in the $n$ blocks case are collectively called $n$ CB algebras.
We would like to understand the structure of these algebras, what are the relations between generators, and whether there are interesting subalgebras or quotients.

The simplest non trivial case is the $n=2$ case.
In this case $X_i$ satisfy a quadratic relation, and it is shown in  ref. \r\Found\ 
 Section 7, that the algebra formed by the $X_i$ is a $A_{N+1}-$Hecke algebra. In this case it is also proven that the ansatz provides a solution to the YBE (1.5) and the commutation (1.4).

An example of an interesting subalgebra is an embedding of the Temperley-Lieb algebra:
We define the operator,
$$E_i=X_i(\lambda),\e$$
where $\lambda$ is the crossing parameter of (1.19). 
Assuming the crossing symmetry (1.24) 
holds, it follows that
$$E\pmatrix{a & b \cr c &d\cr}=\left( {G_b G_c \over G_a G_d}\right)^{1/2} \delta_{a,d},$$
where we denoted $E_i$ above with its explicit indices. From this equation, it follows,
that $E_i$ obeys the Temperley--Lieb algebra (for any $n$),
$$E_i E_{i\pm 1 } E_i=E_i,\qquad E_i^2=b E_i,\qquad E_i E_j=E_j E_i, {\ \ \rm if\ \ } |i-j|>1,\e$$
where
$$b=\prod_{r=0}^{n-2} {\sin(\lambda+\zeta_r)\over \sin(\zeta_r)}.\e$$

{\bf 1.4. The main results}

In this paper we analyze $n$ CB algebras for $n=3,4$ (and $h=v=\bar{h}$).

In the $n=3$ case we prove that the operators $X_i(u)$ constructed by the ansatz satisfy the required relations (1.4),(1.5) and (1.24) 
 if an only if the generators $1,E_i,G_i,G_i^{-1},~i=1,\ldots, N$ where $G_i^{\pm 1}$ are proportional to $X_i,X_i^t,$ form an algebra to which we call the {\it weak Birman-Murakami-Wenzl (BMW)} algebra. This algebra which is defined below, has the property that a simple quotient of it gives the well known BMW algebra \r{\BW,\Mur} .
We conjecture, and have verified in many examples, that the above generators satisfy the BMW algebra itself. We conjecture that most of the BMW algebra relations hold for all $n.$

In the $n=4$ case, assuming this conjecture regarding the BMW relations for general $n,$ we describe an algebra which is a generalization of the BMW algebra over the same set of generators, and which is equivalent to YBE.

We then consider two explicit special cases with $n=4$ blocks. The first is the $G_2$ model and the second is $SU(2)~3\times3$ model.
For the first model we show that by adding two new types of generators $H_i,K_i,~i=1,\ldots, N$ the algebra formed by $1,E_i,G_i,G_i^{-1},H_i,K_i,~i=1,\ldots, N$ is the Kalfagianni-Kuperberg algebra (defined in \r\Kal,
following the work of ref. \r\Kuper).
We then show that by defining $H_i,K_i$ in an analogous way, very similar relations hold in the case of $SU(2)~3\times3$ model. The algebra we find there is new, as far as we know. We extend the definition of the new generators $H_i,K_i$ to any $n=4$ fusion IRF model. In Sections (6,7) we show that the relations we find also extend in a similar manner to the general $n=4$ theories.

\mysec{$n=3$ case}

Consider the case of $n=3$. Auppose the crossing relation
$$
\omega\left(\matrix{a & b \cr c &d\cr} \bigg |  \zeta_0 - u \right) = \left( {G_b G_c \over G_a G_d} \right)^{ 1 \over 2} \omega\left( \matrix{c & a \cr d &b\cr} \bigg | u \right),
\e$$
where $G_a$ is the multiplier.

Set $s_0 = e^{i\zeta_0}$, $s_1 = e^{i\zeta_1}$. 

Recall equation (1.25) and the ansatz. We have
$$
E_i = X_i(\zeta_0) = \frac{\sin(2\zeta_0)\sin(\zeta_0+\zeta_1)}{\sin(\zeta_0)\sin(\zeta_1)} P_i^0 = \frac{\left({s_0}^2 {s_1}^2 - 1\right) \left({s_0}^2 + 1\right)}{\left({s_1}^2 - 1\right) {s_0}^2} P_i^0.
\e$$
Put 
$$\eqalign{
G_i = 4 \sin(\zeta_0)\sin(\zeta_1) e^{-i\zeta_0} X_i & = -e^{-2i\zeta_0-i\zeta_1} P_i^0 + e^{-i\zeta_1} P_i^1 - e^{i\zeta_1} P_i^2 \cr
& = -s_0^{-2}s_1^{-1} P_i^0 + s_1^{-1} P_i^1 - s_1P_i^2.}
\e$$

Then
$$\eqalign{
& P_i^0 = \frac{({s_1}^2 - 1) {s_0}^2}{({s_0}^2 {s_1}^2 - 1) ({s_0}^2 + 1)} E_i, \cr
& P_i^1 = \frac{{s_1}^2 - 1}{({s_0}^2 + 1)  ({s_1}^2 + 1)  {s_0}^2}E_i+\frac{{s_1}^2}{{s_1}^2 + 1}G_i^2+\frac{{s_1}^3}{{s_1}^2 + 1}G_i \cr
& P_i^2 = - \frac{({s_1}^2 - 1)}{({s_0}^2  {s_1}^2 - 1)  ({s_1}^2 + 1)  {s_0}^2  {s_1}^2}E_i- \frac{1}{({s_1}^3 + s_1)}G_i+\frac{1}{{s_1}^2 + 1}G_i^2,}
\e $$
and $G_i E_i = E_i G_i = l^{-1} E_i$, where $l = -s_0^2s_1$.

Using $P_i^0 + P_i^1 + P_i^2 = 1$, we obtain
$$
G_i^2 = - \frac{({s_1}^2 - 1)}{{s_0}^2  {s_1}^2}E_i+1+(\frac{1}{s_1} - s_1)G_i.
\e $$

Denote $m=-2i\sin(\zeta_1) = s_1^{-1} - s_1$. Then 
$$
G_i = - \frac{m}{l}E_i G_i^{-1}+G_i^{-1}+m.
\e $$
This implies the skein relation
$$
m(E_i-1) = G_i^{-1} - G_i.
\e $$

Using the expression for $G_i^2$, we have
$$\eqalign{
& P_i^0 = \frac{({s_1}^2 - 1) {s_0}^2}{({s_0}^2 {s_1}^2 - 1) ({s_0}^2 + 1)} E_i, \cr
& P_i^1 = - \frac{({s_1}^2 - 1)}{({s_0}^2 + 1)  ({s_1}^2 + 1)}E_i+\frac{{s_1}^2}{{s_1}^2 + 1}+\frac{s_1}{{s_1}^2 + 1}G_i \cr
& P_i^2 = - \frac{({s_1}^2 - 1)}{({s_0}^2  {s_1}^2 - 1)  ({s_1}^2 + 1)}E_i- \frac{s_1}{({s_1}^2 + 1)}G_i+\frac{1}{{s_1}^2 + 1}.}
\e $$

Therefore
$$\eqalign{
X_i(u) & = \frac{1}{\left(s_0^2-1\right) \left(s_1^2-1\right)}(q_1 + q_2 e^{2iu} + q_3 e^{-2iu}), \cr
q_1 & = ({s_1}^2 - 1) {s_0}^2 + (1 - {s_1}^2) E_{i} + ({s_0}^2 + 1) s_1 G_{i}, \cr
q_2 & = (1 - {s_1}^2) + ({s_1}^2 - 1) E_{i} - s_1 G_{i}, \cr
q_3 & = - {s_0}^2 s_1 G_{i}. }
\e $$

By equations (1.26-1.27),  $E_iE_{j}E_i=E_i$, when $|i-j|=1,$ and $E_i^2 = bE_i$, where $b = \frac{\sin(2\zeta_0)\sin(\zeta_0+\zeta_1)}{\sin(\zeta_0)\sin(\zeta_1)}=\frac{\left({s_0}^2 {s_1}^2 - 1\right) \left({s_0}^2 + 1\right)}{\left({s_1}^2 - 1\right) {s_0}^2}$.

The skein relation implies that for $|i-j|=1$, 
$$
\eqalign{
& (E_i-1)E_jG_i = \frac{1}{m} (G_i^{-1}-G_i)E_jG_i \cr
& G_i E_j( E_i-1 ) = \frac{1}{m} G_i E_j ( G_i^{-1} - G_i ),
}
\e $$
and 
$$
G_i^{-1}E_jG_i = G_i^{-1}(1+\frac{1}{m}(G_j^{-1}-G_j))G_i, \quad G_jE_iG_j^{-1} = G_j (1+\frac{1}{m}(G_i^{-1}-G_i))G_j^{-1}.
\e $$
Therefore
$$\eqalign{
& E_i E_j G_i = E_j G_i - \frac{1}{m} G_i E_j G_i + \frac{1}{m} G_i^{-1} E_j G_i, \cr
& G_i E_j E_i = G_i E_j + \frac{1}{m} G_i E_j G_i^{-1} - \frac{1}{m} G_i E_j G_i.}
\e $$
and $G_i^{-1}E_jG_i = G_jE_iG_j^{-1}$.

Using $X_i(u)X_{i+1}(u+v)X_i(v) = X_{i+1}(v)X_{i}(u+v)X_{i+1}(u)$ and the above relations, we obtain $19$ equations. The only two independent equations are the following equations.
$$\eqalign{
& {({s_1}^2 - 1)}^2 ({s_0}^2 {s_1}^2 - {s_1}^2 + 1) E_{i}- {({s_1}^2 - 1)}^2 ({s_0}^2 {s_1}^2 - {s_1}^2 + 1) E_{i+1}
\cr
& +{({s_1}^2 - 1)}^2 s_1 E_{i+1} G_{i}+({s_1}^4 - {s_1}^2)G_{i} E_{i+1} G_{i}+{({s_1}^2 - 1)}^2 s_1 E_{i} G_{i+1} E_{i}
\cr 
& +{({s_1}^2 - 1)}^2 s_1 G_{i} E_{i+1} - {({s_1}^2 - 1)}^2 s_1 E_{i} G_{i+1} + ({s_1}^2 - {s_1}^4)G_{i+1} E_{i} G_{i+1} \cr
& - {({s_1}^2 - 1)}^2 s_1 E_{i+1} G_{i} E_{i+1}- {({s_1}^2 - 1)}^2 s_1 G_{i+1} E_{i} = 0,}
\e $$

$$\eqalign{
& - {({s_1}^2 - 1)}^2 ({s_0}^4 + {s_0}^2 {s_1}^2 - {s_1}^2 + 1) E_{i}+{({s_1}^2 - 1)}^2 ({s_0}^4 + {s_0}^2 {s_1}^2 - {s_1}^2 + 1) E_{i+1}
\cr
& - {({s_1}^2 - 1)}^2 ({s_0}^4 + 1) s_1 E_{i+1} G_{i}+{({s_1}^2 - 1)}^2 ({s_0}^4 + 1) s_1G_{i+1}E_{i} \cr
& - ({s_1}^2 - 1) ({s_0}^4 + 1) {s_1}^2G_{i}E_{i+1}G_{i} - ({s_0}^2 + 1) {({s_1}^2 - 1)}^2 s_1E_{i}G_{i+1}E_{i} \cr
& - {({s_1}^2 - 1)}^2 ({s_0}^4 + 1) s_1G_{i}E_{i+1}+{({s_1}^2 - 1)}^2 ({s_0}^4 + 1) s_1E_{i}G_{i+1} 
\cr
& +({s_1}^2 - 1) ({s_0}^4 + 1) {s_1}^2G_{i+1}E_{i}G_{i+1}+({s_0}^2 + 1) {({s_1}^2 - 1)}^2 s_1E_{i+1}G_{i}E_{i+1}=0.}
\e $$

These equations are equivalent to the following:
$$
{s_0}^2 s_1E_{i}- {s_0}^2 s_1E_{i+1}+E_{i}G_{i+1}E_{i}-E_{i+1}G_{i}E_{i+1}=0
\e $$
and
$$\eqalign{
& {({s_1}^2 - 1)}^2E_{i}- {({s_1}^2 - 1)}^2E_{i+1}+(s_1 - {s_1}^3)E_{i+1}G_{i}+({s_1}^2 - 1) s_1G_{i+1}E_{i} \cr
& - {s_1}^2G_{i}E_{i+1}G_{i} +(s_1 - {s_1}^3)G_{i}E_{i+1}+({s_1}^2 - 1) s_1E_{i}G_{i+1}+{s_1}^2G_{i+1}E_{i}G_{i+1}=0.}
\e $$

Thus, assume the crossing relation (1.24), or even only its consequence eq. (1.26). We have proved that $X_i(u)$ satisfies the Yang-Baxter equation 
$$
X_i(u) X_{i+1}(u+v) X_i(v) = X_{i+1}(v) X_i(u+v) X_i(u),
\e $$
if and only if $G_i, E_i$ satisfies the following relations:
$$
\eqalign{
& G_i G_j = G_j G_i, \quad |i-j| \ge 2, \cr
& G_i G_j G_i = G_j G_i G_j, \quad |i-j|=1, \cr
& m(E_i-1) = G_i^{-1}-G_i, \cr
& G_i E_i = l^{-1} E_i, \cr
& E_i G_j E_i - l E_i - E_j G_i E_j + lE_j =0, \quad |i-j|=1, \cr
& m^2(E_i-E_j)+m(E_jG_i-G_jE_i +G_iE_j-E_iG_j)-G_iE_jG_i +G_jE_iG_j=0, \quad |i-j|=1,}
\e $$
where $m = \frac{l-l^{-1}}{b-1} = s_1^{-1}-s_1$, $l=-s_0^2s_1$.

We call the above relations weak BMW relation and we call the algebra generated by $G_i, E_i$ subject to the weak BMW relation the weak BMW algebra.

In fact, for general $n$ blocks, using the ansatz, the Yang-Baxter equation implies that 
$$
\eqalign{
& G_i G_j = G_j G_i, \quad |i-j| \ge 2, \cr
& G_i G_j G_i = G_j G_i G_j, \quad |i-j|=1, \cr
& G_i E_i = l^{-1} E_i,}
\e $$
for some $l$.

%

\def\center#1\endcenter{\centerline{#1}}
\def\threeline#1#2#3{\displaylines{\qquad#1\hfill\llap{(\adveq\myeq)}\cr\hfill#2\hfill \cr
\hfill#3\qquad\cr}}

\def\frac#1#2{{#1\over #2}}

\mysec{The 4--CB algebra.}
 We focus now on the four blocks case, $n=4$.

We find it convenient, for future use, to scale the braiding matrices as follows. We have
$$X_i=\lim_{u\rarrow i\infty} \exp(3i u) X_i(u),\qquad X_i^t=\lim_{u\rarrow -i\infty} \exp(-3i u) X_i(u).\e$$
We define the operators $G_i$, $G_i^{-1}$ and $E_i$ by,
$$\manyeq{
G_i&=8 e^{-3i \zeta_0/2} \sin(\zeta_0)\sin(\zeta_1)\sin(\zeta_2) X_i,\cr
G_i^{-1}&=8 e^{3i \zeta_0/2} \sin(\zeta_0)\sin(\zeta_1)\sin(\zeta_2) X_i^t,\cr
E_i&=X_i(\zeta_0).\cr}$$
The normalization is taken so that $G_i^{-1}$ will be the inverse of $G_i$,
$$G_i G_i^{-1}=1_i,\e$$
in view of the inversion relation eq. (1.22). 

We can now express the projection operators $P_i^a$ in terms of $G_i$, $G_i^{-1}$ and $E_i$.
This is given by solving the set of equations which are obtained from the ansatz for the 
Boltzmann weights, eqs. (1.20--1.21),
\def\frac#1#2{{#1\over #2}}
 $$
\twoline{
G_i=i e^{-\frac{5}{2} i \zeta_0-i \zeta_1-i \zeta_2} \big(e^{2 i \zeta_0} P_i^1-e^{2 i \zeta_0+2 i \zeta_1} P_i^2+}{e^{2 i \zeta_0+2 i \zeta_1
+2 i \zeta_2} P_i^3
-P_i^0\big),}
 $$
 $$
\twoline{
G^{-1}_i=i e^{\frac{1}{2} i \zeta_0-i \zeta_1-i \zeta_2} \big(e^{2 i \zeta_0
+2 i \zeta_1+2 i \zeta_2} P_i^0-}{e^{2 i \zeta_1+2 i \zeta_2} \
P_i^1+e^{2 i \zeta_2} P_i^2-P_i^3\big),}
$$
 $$
 \twoline{ E_i=
\frac{e^{-3 i \zeta_0} \left(1+e^{2 i \zeta_0}\right) \left(-1+e^{i \zeta_0+i \zeta_1}\right) \left(1+e^{i \zeta_0+i \zeta_1}\right)}{\left(-1+e^{i \zeta_1}\right) \left(1+e^{i \zeta_1}\right) \left(-1+e^{i \zeta_2}\right) \left(1+e^{i \zeta_2}\right)} \times}{
\left(-1+e^{i \zeta_0+i \zeta_2}\right)  \left(1+e^{i \zeta_0+i \zeta_2}\right) P_i^0,}
 $$
 along with
 $$\sum_{a=0}^3 P^a_i=1.\e$$

 Our purpose is to describe the algebra obeyed by $G_i$, $G_i^{-1}$ and $E_i$. These are defined 
 by
 $$\manyeq{G_i&=8\left[ \prod_{r=0}^2 \sin(\zeta_r)\right] e^{-3i\lambda/2} X_i,\cr
 G_i^{-1}&=8\left[ \prod_{r=0}^2 \sin(\zeta_r)\right] e^{3i\lambda/2} X_i^t,\cr
 E_i&=X_i(\lambda).\cr}$$
By slight abusing the notations, we also call this algebra 4--CB (conformal braiding) algebra.

 Due to the inversion relation, eq. (1.22) we have the relation,
 $$G_i G_i^{-1}=1_i.\e$$
 The phase in eq. (3.8) is arbitrary and is set to simplify the 4--CB algebra. We have also the
 braiding relation,
 $$G_i G_{i+1} G_i=G_{i+1} G_i G_{i+1},\qquad G_i G_j=G_j G_i {\ \rm if \ }|i-j|\geq 2.\e$$
 
 We already know that $E_i$ obeys the Temperley--Lieb algebra, eqs. (1.26--1.27),
 $$E_i E_{i\pm1} E_i=E_i,\qquad E_i^2=b E_i,\qquad{\rm where\ \ } b=\prod_{r=0}^2 {\sin(\lambda+\zeta_r)\over \sin(\zeta_r)}.\e$$
 
The next relations are,
$$G_i E_i=E_i G_i=l^{-1} E_i,\e$$
which follow by taking the ansatz eqs. (1.20--1.21) for $G_i$ and using the projection relations, eq. (1.18),
$P_i^a P_i^0=P_i^0 P_i^a=\delta_{a,0} P_i^0$. The value of $l$ is,
$$l=i e^{i(3\lambda/2+\zeta_0+\zeta_1+\zeta_2)}.\e$$

We can calculate,
 $$
\twoline{G_i^2=
-e^{-5 i \zeta_0-2 i \zeta_1-2 i \zeta_2} P_i^0-e^{-i \zeta_0-2 i \
\zeta_1-2 i \zeta_2} P_i^1-}{e^{-i \zeta_0+2 i \zeta_1-2 i \zeta_2} \
P_i^2-e^{-i \zeta_0+2 i \zeta_1+2 i \zeta_2} P_i^3.}
$$
We can substitute the expressions for $P_i^a$ from $G_i$, $G_i^{-1}$ and $E_i$, eqs. (3.4--3.6). We 
then find the relation expressing $G_i^2$,
$$\threeline{
G_i^2= i e^{-\frac{1}{2} i \zeta_0-i \zeta_1-i \zeta_2} \left(1-e^{2 i \zeta_1}+e^{2 i \zeta_1+2 i \zeta_2}\right) \
G_i+i e^{-\frac{3}{2} i \zeta_0+i \zeta_1-i \zeta_2} \
G_i^{-1}}{+\frac{e^{-2 i \zeta_0-2 i \zeta_1-2 i \zeta_2} \left(e^{2 i 
\zeta_1}-1\right) \left(1+e^{2 i \zeta_0+2 
i \zeta_1+2 i \zeta_2}\right) \left(e^{2 i \zeta_2}-1\right) 
 }{\left(e^{2 i \zeta_0+ 2 i \zeta_2}-1\right) }E_i}{-e^{-i \zeta_0-2 i 
\zeta_2} \left(1-e^{2 i \zeta_2}+e^{2 i \zeta_1+2 i \zeta_2}\right).}$$
We define the coefficients $\alpha$, $\beta$, $\gamma$ and $\delta$ by writing this equation as
$$G_i^2=\alpha+\beta E_i+\gamma  G_i+\delta G_i^{-1}.\e$$
This is the skein relation.
 
 From the skein relation, eq. (3.16), we find the relation,
 $$G_{i\pm1} G_i E_{i\pm1}=E_i G_{i\pm1} G_i,\e$$
 which follows by expressing $E_i=(G_i^2-\alpha-\gamma G_i-\delta G_i^{-1})/\beta$
 from the skein relation eq. (3.16) and inserting it into eq. (3.10). 
 \par
 The additional relations areconjectural, but hold in all examples we have checked. These relations assume the same form as the Birman--Murakami--Wenzl (BMW) algebra 
\r{\BW,\Mur}\
and are summarized below.
$$\manyeq{
G_{i\pm1} G_i E_{i\pm1}&=E_i E_{i\pm1},\qquad G_{i\pm1} E_i G_{i\pm1}=G_i^{-1} E_{i\pm1} G_i^{-1},\cr
G_{i\pm1} E_i E_{i\pm1}&=G_i^{-1}  E_{i\pm1},\qquad E_{i\pm1} E_i G_{i\pm1}=E_{i\pm1} G_i^{-1},\cr
E _iG_{i\pm1} E_i&=l E_i.\cr}$$

The last relation follows from the Yang Baxter equation, see Section (1). It is 
$$g(i,i+1,i)=g(i+1,i,i+1),\e$$
where 
$$\threeline{
g=a_{1,2,4}+a_{1,3,1}+a_{4,2,1}+i q^{-\zeta_0/2+\zeta_1-\zeta_2} (a_{1,3,4}+a_{4,2,4}+a_{4,3,1})+}{
i q^{\zeta_0/2-\zeta_1+\zeta_2}  (a_{2,3,4}+a_{4,1,4}+a_{4,3,2})+}{
i {q^{\zeta_1+\zeta_2}\over (q^{2\zeta_1}-1)(q^{2\zeta_2}-1)}\left(q^{\zeta_0/2} a_{1,2,1}+q^{-\zeta_0/2} a_{2,1,2} \right)+z a_{4,3,4},}$$
where 
$$\twoline{z={q^{-\zeta_0-2\zeta_1-2\zeta_2} (q^{2\zeta_1}-1)(q^{2\zeta_2}-1)\over q^{2\zeta_0+2\zeta_2-1}}\times}{
\left( 2 q^{2\zeta_0+2\zeta_2} +2 q^{2\zeta_0+2\zeta_1+2\zeta_2}+q^{4\zeta_0+2\zeta_1+4 \zeta_2}+1\right).}$$  
We denoted by $a_{i,j,k} (r,s,t)$ the element of the algebra $a_i[r] a_j[s] a_k[t]$ where
$a_i[r]$ is $G_r, G_r^{-1},E_r$ or $1_r$, if $i=1,2,3,4$, respectively.

This summarizes all the relations of the 4--CB algebra. As we will show below, these are all the relations that follow from the YBE and are equivalent to it. We checked these relations for the model $SU(2)$ fused $3\times 3$ as described in ref.
\r\Second. We also checked these relations for the $G_2$ model
described in Section (5) for $q=0.7$. We find a complete agreement with the 4--CB algebra.
In Section (5) we will prove this algebra for $G_2$.

{\bf 3.1. $n=4$ YBE and relations (3.11)-(3.21):}
Our goal now is to verify that all the relations (3.11)--(3.21),  represent a certain solution of the YBE, if(f) the parameters  are fixed in appropriate way. The idea of the check is rather simple and follows closely the analogous  consideration for 3-blocks case,  described in the previous section.

Denoting $s_0 = e^{i\zeta_0}$, $s_1 = e^{i\zeta_1}$, $s_2 = e^{i\zeta_2}$, we get the projectors
$$\eqalign{
& P_i^0 =\frac{s_0^3 \left(s_1^2-1\right) \left(s_2^2-1\right)}{\left(s_0^2+1\right) \left(s_0^2 s_1^2-1\right)  \left(s_0^2 s_2^2-1\right)} E_i, \cr
& P_i^1 = \frac{i s_2 s_1^3 }{\sqrt{s_0} \left(s_1^2+1\right) \left(s_1^2 s_2^2-1\right)}G_i^{-1}+\frac{i \sqrt{s_0} s_2 s_1 }{\left(s_1^2+1\right) \left(s_1^2 s_2^2-1\right)} G_i+\cr
& \,\,\,\,\,\,\,\,+\frac{s_0 \left(s_1^2-1\right) \left(s_2^2-1\right) \left(s_0^2 s_1^2 s_2^2+1\right)}{\left(s_0^2+1\right) \left(s_1^2+1\right) \left(s_0^2 s_2^2-1\right)  \left(s_1^2 s_2^2-1\right) }E_i+ \frac{\left(s_2^2-1\right) s_1^2 }{\left(s_1^2+1\right) \left(s_1^2 s_2^2-1\right) } 1_i,\cr
& P_i^2 = \frac{i \sqrt{s_0} s_1 s_2 }{\left(s_1^2+1\right) \left(s_2^2+1\right)}G_i-\frac{i s_1 s_2}{\sqrt{s_0} \left(s_1^2+1\right) \left(s_2^2+1\right)}G_i^{-1}-\cr
&\,\,\,\,\,\,\,\,-\frac{s_0 \left(s_1^2-1\right) \left(s_2^2-1\right)  \left(s_0^2 s_1^2 s_2^2+1\right)}{\left(s_0^2 s_1^2-1\right) \left(s_1^2+1\right) \left(s_0^2 s_2^2-1\right) \left(s_2^2+1\right)}E_i+\frac{\left(s_1^2 s_2^2+1\right) }{\left(s_1^2+1\right) \left(s_2^2+1\right)}1_i,}
\e $$ 
where (3.7) is used in order to eliminate $P_i^3$. Note that the skein relation (3.16) allows also to use other choices of any three independent parameters.

Taking into account  the explicit  form of the conjectured trigonometric solution eqs. (1.20--1.21), the  Yang Baxter equation (1.5) can be written in the form
$$
YBE\Rightarrow\,\,\,\sum_{r,s} C_{rs}^{YBE} q_1^r q_2^s=0,\e
$$
where $q_{1}=e^{i u}$, $q_{2}=e^{i v}$,  the integers $r,s$ are in the region $[0,2n]$, and the coefficients $C_{rs}^{YBE}$ are expressed in terms of the generators $E_i$, $G_i$, $G_i^{-1}$ and $1_i$. 
Hence, in order to fulfil the YBE for generic values of the spectral parameters $u,v$ one has to ensure that the coefficients $C_{rs}^{YBE}=0$, for all $r$ and $s$.

In the  $4$-blocks case simple counting shows that there are  $37$ relations, $r\in [0,8]$ step $2$, and $s\in [Max(r-6,0),Min(r+6,8)]$ step $2$. Taking into account the factorised form of the coefficients $f_a(u)$ in terms of sine functions, see eq.(1.21), we note that for general $n$  the number of the relations is given by Hex number,  i.~e. by the number of partitions of $6(n-1)$ into at most 3 parts, which is $N(n)=3(n-1)n+1$, so that for lower values of $n$ we have $1,7,19,37,61,91,...$. 

In our case a few relations are written below explicitly 
$$\eqalign{
&(r,s)=(0,0):\,\,\,\,a_{1,1,1}(i,i+1,i)-a_{1,1,1}(i+1,i,i+1)=0,\cr
&(r,s)=(0,2):\,\,\,\,s_1^2 s_2^2 s_0^{3/2} a_{4,1,1}(i+1,i,i+1)-s_2^2 s_0^{3/2} a_{4,1,1}(i+1,i,i+1)\cr
&+s_2^2 s_0^{3/2} a_{1,1,4}(i,i+1,i)+s_1^2 s_0^{3/2} a_{1,1,4}(i,i+1,i)-s_1^2 s_0^{3/2} a_{4,1,1}(i+1,i,i+1)\cr
&-s_1^2 s_2^2 s_0^{3/2} a_{1,1,4}(i,i+1,i)-s_0^{3/2} a_{1,1,4}(i,i+1,i)+s_0^{3/2} a_{4,1,1}(i+1,i,i+1)\cr
&-i s_1 s_2 s_0^2 a_{1,1,1}(i,i+1,i)+i s_1 s_2 s_0^2 a_{1,1,1}(i+1,i,i+1)+i s_1 s_2 s_0 a_{1,1,2}(i,i+1,i)\cr
&-i s_1 s_2 s_0 a_{2,1,1}(i+1,i,i+1)-s_1^2 \sqrt{s_0} a_{1,1,3}(i,i+1,i)+s_1^2 s_2^2 \sqrt{s_0} a_{1,1,3}(i,i+1,i)\cr
&-s_2^2 \sqrt{s_0} a_{1,1,3}(i,i+1,i)+s_1^2 \sqrt{s_0} a_{3,1,1}(i+1,i,i+1)-\sqrt{s_0} a_{3,1,1}(i+1,i,i+1)\cr
&-i s_1 s_2 a_{1,1,1}(i,i+1,i)+i s_1 s_2 a_{1,1,1}(i+1,i,i+1)+\sqrt{s_0} a_{1,1,3}(i,i+1,i)\cr
&-s_1^2 s_2^2 \sqrt{s_0} a_{3,1,1}(i+1,i,i+1)+s_2^2 \sqrt{s_0} a_{3,1,1}(i+1,i,i+1)=0,\cr
&(r,s)=(0,4):\,\,\,\,s_1^2 s_2^2 s_0^{3/2} a_{3,1,1}(i+1,i,i+1)-s_2^2 s_0^{3/2} a_{3,1,1}(i+1,i,i+1)\cr
&+s_2^2 s_0^{3/2} a_{1,1,3}(i,i+1,i)-s_0^{3/2} a_{1,1,3}(i,i+1,i)-s_1^2 s_0^{3/2} a_{3,1,1}(i+1,i,i+1)\cr
&+s_0^{3/2} a_{3,1,1}(i+1,i,i+1)+s_1^2 s_0^{3/2} a_{1,1,3}(i,i+1,i)-s_1^2 s_2^2 s_0^{3/2} a_{1,1,3}(i,i+1,i)\cr
&-i s_1 s_2 s_0^2 a_{1,1,2}(i,i+1,i)+i s_1 s_2 s_0^2 a_{2,1,1}(i+1,i,i+1)+i s_1 s_2 s_0 a_{1,1,1}(i,i+1,i)\cr
&-i s_1 s_2 s_0 a_{1,1,1}(i+1,i,i+1)-s_1^2 \sqrt{s_0} a_{1,1,4}(i,i+1,i)+s_1^2 s_2^2 \sqrt{s_0} a_{1,1,4}(i,i+1,i)\cr
&-s_2^2 \sqrt{s_0} a_{1,1,4}(i,i+1,i)+\sqrt{s_0} a_{1,1,4}(i,i+1,i)+s_1^2 \sqrt{s_0} a_{4,1,1}(i+1,i,i+1)\cr
&-i s_1 s_2 a_{1,1,2}(i,i+1,i)+i s_1 s_2 a_{2,1,1}(i+1,i,i+1)-\sqrt{s_0} a_{4,1,1}(i+1,i,i+1)\cr
&-s_1^2 s_2^2 \sqrt{s_0} a_{4,1,1}(i+1,i,i+1)+s_2^2 \sqrt{s_0} a_{4,1,1}(i+1,i,i+1)=0,\cr
&(r,s)=(0,6):\,\,\,\,a_{1,1,2}(i,i+1,i)-a_{2,1,1}(i+1,i,i+1)=0,\cr
&etc.
}\e $$ 
Here we use the notation introduced just below eq.(3.21). We do not write all the relations since their expressions are bulky in general, while the computation is rather straightforward.

Our next task is to implement the relations  (3.11)--(3.19). We perform this gradually. First, we use the ``simple'' parameter-free relations, like the  braiding relations (3.10), parameter-free Temperley--Lieb algebra relations (3.11), the relation (3.18), and all the relations, which follow from these relations. We chose this order not to end up with some particular solution of the YBE, obeying softer algebraic constraints, but existing only for special choices of the parameters. The relations are  
$$\eqalign{
&a_{1,1,1}(i+1,i,i+1)= a_{1,1,1}(i,i+1,i),\,\,a_{2,1,1}(i+1,i,i+1)= a_{1,1,2}(i,i+1,i),\cr
&a_{1,1,2}(i+1,i,i+1)= a_{2,1,1}(i,i+1,i),\,\,a_{2,2,1}(i+1,i,i+1)= a_{1,2,2}(i,i+1,i),\cr
&a_{1,2,2}(i+1,i,i+1)= a_{2,2,1}(i,i+1,i),\,\,a_{2,2,2}(i+1,i,i+1)= a_{2,2,2}(i,i+1,i),\cr
&a_{3,1,1}(i+1,i,i+1)= a_{1,1,3}(i,i+1,i),\,\,a_{1,1,3}(i+1,i,i+1)= a_{3,1,1}(i,i+1,i),\cr
&a_{2,3,1}(i+1,i,i+1)= a_{1,3,2}(i,i+1,i),\,\,a_{2,4,1}(i+1,i,i+1)= a_{1,4,2}(i,i+1,i),\cr
&a_{1,3,2}(i+1,i,i+1)= a_{2,3,1}(i,i+1,i),\,\,a_{1,4,2}(i+1,i,i+1)= a_{2,4,1}(i,i+1,i),\cr
&a_{2,2,3}(i+1,i,i+1)= a_{3,2,2}(i,i+1,i),\,\,a_{3,2,2}(i+1,i,i+1)= a_{2,2,3}(i,i+1,i).
}
\e $$ 
For example, to obtain the last relation $a_{3,2,2}(i+1,i,i+1)= a_{2,2,3}(i,i+1,i)$, which is explicitly $E_{i+1}G_{i}^{-1}G_{i+1}^{-1}=G_{i}^{-1}G_{i+1}^{-1}E_{i}$, 
we multiply the equation (3.17) from both sides  by $G_{i}^{-1}G_{i+1}^{-1}$  and then use the inversion relation  (3.9).

Now we are in the position to implement the relations involving parameters (like skein relation, etc.) and also all their derivatives. 
In principle, this can be done in deferent ways, our guiding principle is to exclude all linear dependent  elements, simultaneously choosing among  the linear independent elements those with the maximal numbers of $1_i$ generators. First set of the relations is
$$\eqalign{
&a_{1,2,3}(x)= \alpha  a_{4,3,3}(x)+\beta  a_{3,3,3}(x)+\gamma  a_{1,3,3}(x)+\delta  a_{2,3,3}(x),\cr
&a_{2,1,3}(x)= -\frac{\beta l a_{3,3,3}(x)}{\delta }-\frac{\alpha  a_{2,3,3}(x)}{\delta }-\frac{\gamma  a_{4,3,3}(x)}{\delta }+\frac{a_{1,3,3}(x)}{\delta },\cr
&a_{3,2,1}(x)= \alpha  a_{3,3,4}(x)+\beta  a_{3,3,3}(x)+\gamma  a_{3,3,1}(x)+\delta  a_{3,3,2}(x),\cr
&a_{3,1,2}(x)= -\frac{\beta  l a_{3,3,3}(x)}{\delta }-\frac{\alpha  a_{3,3,2}(x)}{\delta }-\frac{\gamma  a_{3,3,4}(x)}{\delta }+\frac{a_{3,3,1}(x)}{\delta },
}
\e $$ 
which follow from the skein relation. Here $x$ stands for either $(i+1,i,i+1)$ or $(i,i+1,i)$. For example, in order to obtain the first one we use the 
relation $G_{i+1}^{-1}E_i  =G_i E_{i+1}E_i$  from eq. (3.18), and then the 4-blocks skein relation to exclude $G_i^2$ on the first place. 

Similarly, we get
$$\eqalign{
&a_{3,1,3}(i,i+1,i)= l a_{4,3,4}(i+1,i,i+1),\,\,a_{3,1,3}(i+1,i,i+1)= l a_{4,3,4}(i,i+1,i),\cr
&a_{2,4,3}(i+1,i,i+1)= l a_{4,3,4}(i,i+1,i),\,\,a_{2,4,3}(i,i+1,i)= l a_{4,3,4}(i+1,i,i+1),\cr
&a_{3,4,2}(i+1,i,i+1)= l a_{4,3,4}(i,i+1,i),\,\,a_{3,4,2}(i,i+1,i)= l a_{4,3,4}(i+1,i,i+1),\cr
&a_{3,2,3}(i,i+1,i)= \frac{a_{4,3,4}(i+1,i,i+1)}{l},\,\,a_{3,2,3}(i+1,i,i+1)= \frac{a_{4,3,4}(i,i+1,i)}{l}\cr
&a_{1,4,3}(i+1,i,i+1)= \frac{a_{4,3,4}(i,i+1,i)}{l},\,\,a_{1,4,3}(i,i+1,i)= \frac{a_{4,3,4}(i+1,i,i+1)}{l},\cr
&a_{3,4,1}(i+1,i,i+1)= \frac{a_{4,3,4}(i,i+1,i)}{l},\,\,a_{3,4,1}(i,i+1,i)= \frac{a_{4,3,4}(i+1,i,i+1)}{l}.
}
\e $$ 
In particular, the first two correspond to the last equation in (3.18) and others can be derived from it. 

We also have 
$$\eqalign{
&a_{1,4,1}(i\pm1,i,i\pm1)=\cr
&\,\,\,\,\,\, \alpha  a_{4,4,4}(i,i\pm1,i)+\beta  a_{4,3,4}(i,i\pm1,i)+\gamma  a_{4,1,4}(i,i\pm1,i)+\delta  a_{4,2,4}(i,i\pm1,i),\cr
&a_{2,4,2}(i\pm1,i,i\pm1)=\cr
&\,\,\,\,\,\, \frac{a_{4,1,4}(i,i\pm1,i)}{\delta }-\frac{\beta  l a_{4,3,4}(i,i\pm1,i)}{\delta }-\frac{\alpha  a_{4,2,4}(i,i\pm1,i)}{\delta }-\frac{\gamma  a_{4,4,4}(i,i\pm1,i)}{\delta },
}
\e $$ 
and
$$\eqalign{
a_{3,4,3}(i\pm1,i,i\pm1)=  \left(\frac{1}{\beta l^2}-\frac{\alpha}{\beta} -\frac{\gamma }{\beta l}-\frac{\delta  l}{\beta}\right) a_{4,3,4}(i,i\pm1,i),
}
\e $$ 
which follow again from the skein relation, and from the relation (3.12).

And also some trivial relations involving unity operator, like 
$$\eqalign{
&a_{2,4,1}(i,i+1,i)=a_{2,4,1}(i+1,i,i+1),\cr
&a_{1,4,2}(i,i+1,i)= a_{2,4,1}(i,i+1,i),\cr
&a_{1,4,2}(i+1,i,i+1)= a_{2,4,1}(i+1,i,i+1),\cr
&a_{k,s,4}(i+1,i,i+1)= a_{4,k,s}(i,i+1,i),\cr
&a_{4,k,s}(i+1,i,i+1)= a_{k,s,4}(i,i+1,i),\cr
&a_{k,4,4}(i,i+1,i)= a_{4,k,4}(i+1,i,i+1),\cr
&a_{4,4,k}(i,i+1,i)= a_{4,k,4}(i+1,i,i+1),\cr
&a_{k,4,4}(i+1,i,i+1)= a_{4,k,4}(i,i+1,i),\cr
&a_{4,4,k}(i+1,i,i+1)= a_{4,k,4}(i,i+1,i),\cr
}
\e $$
which are valid for any $k,s$. For example, the first relation is  $1_i 1_{i+1}=1_{i+1} 1_i$, which is obviously true.

Finally, we use 
$$\eqalign{
&l= i q^{\frac{5 \zeta_0}{2}+\zeta_1+\zeta_2},\cr
&\alpha = -q^{-\zeta_0-2 \zeta_2} \left(-q^{2 \zeta_2}+q^{2 \zeta_1+2 \zeta_2}+1\right),\cr
&\beta = \frac{q^{-2 \zeta_0-2 \zeta_1-2 \zeta_2} \left(q^{2 \zeta_1}-1\right) \left(q^{2 \zeta_2}-1\right) \left(q^{2 \zeta_0+2 \zeta_1+2 \zeta_2}+1\right)}{q^{2 \zeta_0+2 \zeta_2}-1},\cr
&\gamma = i q^{-\frac{\zeta_0}{2}-\zeta_1-\zeta_2} \left(-q^{2 \zeta_1}+q^{2 \zeta_1+2 \zeta_2}+1\right),\cr
&\delta = i q^{-\frac{3}{2} \zeta_0+\zeta_1-\zeta_2},
}
\e $$ 
or in terms of $s_1,s_2,s_3$,
$$\eqalign{
&l=i s_0^{5/2} s_1 s_2,\cr
&\alpha=-\frac{s_1^2 s_2^2-s_2^2+1}{s_0 s_2^2},\cr
&\beta =\frac{\left(s_1^2-1\right)  \left(s_2^2-1\right) \left(s_0^2 s_1^2 s_2^2+1\right)}{s_0^2 s_1^2 s_2^2 \left(s_0^2 s_2^2-1\right)},\cr
&\gamma = \frac{i \left(s_2^2 s_1^2-s_1^2+1\right)}{\sqrt{s_0} s_1 s_2},\cr
&\delta =\frac{i s_1}{s_0^{3/2} s_2}.
}
\e $$

Substituting the relations (3.14)-(3.21) into 37 relations obtained from the YBE, eq.(3.24), we find that some of the YBE relations become fulfilled, and we are left with 19 relations. 
For these remaining relations we find that they are compatible if and only if the relation $g(i,i+1,i)= g(i+1,i,i+1)$, eqs. (3.19--3.20), is imposed.

\mysec{Three parameter link invariant.}
Let $BMW'_n$ be the 4-CB algebra described in Section (3) with generators $1,G_1,...,G_{n-1}$ and also $E_i$, $i=1,2,\dots,n-1$.
Let $b$ be as in eq. (3.11) and $l$ as in eq. (3.13).
We conjecture the existence of a unique trace function
$$\tau:\bigcup BMW'_n\to \bf{C},$$ which satisfies the following properties

1)$\tau(a+b)=\tau(a)+\tau(b).$

2) $\tau(ab)=\tau(ba).$

3)$\tau(1)=1$ and $\tau(E_i)=b.$

4) $\tau(w\cdot G_n)=l^{-1}\tau(w),$ and $\tau(w\cdot G_n^{-1})=l\tau(w)$, where $w\in BMW'_n.$

The existence of such trace will allow us to define a link invariant extending the results of ref.
\r\Second. We do it in the following way. We assign $E_i$ and $G_i$ to diagrams, as in fig. (1). 
Given a composite diagram, we assign it an expression in terms of $E_i$ and  $G_i$ in the expected way, and apply $\tau$ to the result. We denote by $L(D)$ the result of this assignment, performed on a diagram $D.$ 

\vskip 1cm
\center
\tikzpicture[scale=1]
\draw (34.2,.3) node {$G_i\,\,\longmapsto$};
\draw [line width=1pt] (35.4,-.5) -- (35.4,1.2);
\draw (35.8,.3) node {$\dots$};
\draw [line width=1pt] (36.2,-.5) -- (36.2,1.2);
\draw [line width=1pt] (37.,-.5) -- (37.8,1.2);
\draw [line width=1pt] (37.,1.2) -- (37.33,.5);
\draw [line width=1pt] (37.48,.2) -- (37.8,.-.5);
\draw [line width=1pt] (38.6,-.5) -- (38.6,1.2);
\draw (39.,.3) node {$\dots$};
\draw [line width=1pt] (39.4,-.5) -- (39.4,1.2);
\draw (35.4,1.5) node {1};
\draw (36.2,1.5) node {i-1};
\draw (37.,1.5) node {i};
\draw (37.8,1.5) node {i+1};
\draw (38.6,1.5) node {i+2};
\draw (39.4,1.5) node {n};
\draw (40.5,.3) node {and};
\draw (42.2,.3) node {$E_i\,\,\longmapsto$};
\draw [line width=1pt] (43.4,-.5) -- (43.4,1.2);
\draw (43.8,.3) node {$\dots$};
\draw [line width=1pt] (44.2,-.5) -- (44.2,1.2);
\draw [line width=1pt] (46.6,-.5) -- (46.6,1.2);
\draw (47.,.3) node {$\dots$};
\draw [line width=1pt] (47.4,-.5) -- (47.4,1.2);
\draw (43.4,1.5) node {1};
\draw (44.2,1.5) node {i-1};
\draw (45.,1.5) node {i};
\draw (45.8,1.5) node {i+1};
\draw (46.6,1.5) node {i+2};
\draw (47.4,1.5) node {n};
\draw  [line width=1pt] (45.,1.2) to[out=-80,in=180] node [sloped,above] {} (45.4,.65);
\draw  [line width=1pt] (45.4,.65) to[out=0,in=-100] node [sloped,above] {} (45.8,1.2);
\draw  [line width=1pt] (45.,-.5) to[out=80,in=180] node [sloped,above] {} (45.4,.1);
\draw  [line width=1pt] (45.4,.1) to[out=0,in=100] node [sloped,above] {} (45.8,-.5);
\endtikzpicture
\endcenter
\vskip 5mm
\center
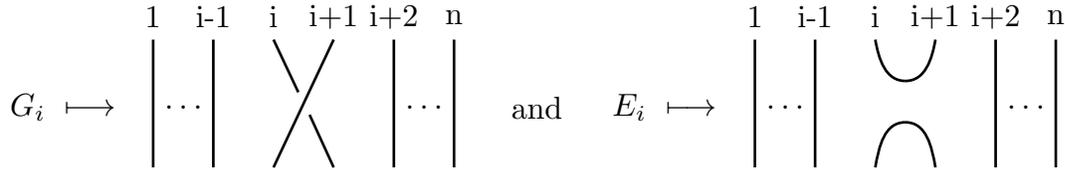
\Fig 1. {The isomorphism between 4--CB and the tangle algebra.}
\endcenter
\vskip5mm 

From the properties of $\tau$ and the relations of $BMW'_n$ one can show that $L$ has the following properties:

1) $L(O)=b,$

2) $L(S_r)=l^{-1} L(S)$ and $L(S_l)=l L(S)$,

3) $L$ is unchanged by Reidemeister moves II,III, which are described in fig (2).

Here $O$ is the standard diagram for the unknot, $S$ is a strand and $S_r$ (respectively $S_l$) is
the same strand with a right handed curl (respectively left handed) as in type I Reidemeister move. 
The third property follows from the relations of the 4-CB algebra.

\vskip 1cm
\center
\tikzpicture[scale=1]
\draw [line width=1pt] (32,0) -- (32,1.2);
\draw [line width=1pt] (32,0) -- (33.5,0);
\draw [line width=1pt] (33.5,0) -- (33.5,.8);
\draw [line width=1pt] (32.8,.8) -- (33.5,.8);
\draw [line width=1pt] (32.8,.8) -- (32.8,.1);
\draw [line width=1pt] (32.8,-.1) -- (32.8,-.5);
\draw [line width=1pt] (34.8,-.5) -- (34.8,1.2);
\draw (34.2,.3) node {$\rightarrow$};
\draw (33.5,1.9) node {Type I};
\draw [line width=1pt] (37,0) -- (37,-.5);
\draw [line width=1pt] (37,1.2) -- (37,.8);
\draw [line width=1pt] (37,.8) -- (37.7,.8);
\draw [line width=1pt] (37,0) -- (37.7,0);
\draw [line width=1pt] (37.9,.8) -- (38.5,.8);
\draw [line width=1pt] (37.9,0) -- (38.5,0);
\draw [line width=1pt] (38.5,0) -- (38.5,.8);
\draw [line width=1pt] (37.8,-.5) -- (37.8,1.2);
\draw [line width=1pt] (39.8,-.5) -- (39.8,1.2);
\draw [line width=1pt] (40.4,-.5) -- (40.4,1.2);
\draw (39.2,.3) node {$\rightarrow$};
\draw (38.8,1.9) node {Type II};
\draw [line width=1pt] (42,0) -- (42,-.5);
\draw [line width=1pt] (42,0) -- (43.,0);
\draw [line width=1pt] (43.,0) -- (43.,1.2);
\draw [line width=1pt] (42.,.8) -- (42.9,.8);
\draw [line width=1pt] (43.1,.8) -- (43.5,.8);
\draw [line width=1pt] (42.5,.1) -- (42.5,.7);
\draw [line width=1pt] (42.5,.9) -- (42.5,1.2);
\draw [line width=1pt] (42.5,-.1) -- (42.5,-.5);
\draw [line width=1pt] (44.8,0) -- (45.2,0);
\draw [line width=1pt] (45.4,0) -- (46.3,0);
\draw [line width=1pt] (45.3,-.5) -- (45.3,.8);
\draw [line width=1pt] (45.3,.8) -- (46.3,.8);
\draw [line width=1pt] (46.3,.8) -- (46.3,1.2);
\draw [line width=1pt] (45.8,.1) -- (45.8,.7);
\draw [line width=1pt] (45.8,.9) -- (45.8,1.2);
\draw [line width=1pt] (45.8,-.1) -- (45.8,-.5);
\draw (44.2,.3) node {$\rightarrow$};
\draw (44.3,1.9) node {Type III};
\endtikzpicture
\endcenter
\vskip 5mm
\center
\Fig 1. {Reidemeister moves.}
\endcenter
\vskip 5mm

We correct $L$ to form a Markov trace by putting 
$$\nu(K)=l^{\omega(K)} L(K),\e$$
where $K$ is link, $\nu(K)$ is the link invariant, $\omega(K)$ is the writhe of the link defined as the
number of left crossings minus the number of right crossings. By properties 2),3) above we see that $\nu$ is a link invariant.

If the conjecture is correct then the resulting link invariant belongs to a three parameter family of invariants, where the parameters are the $\zeta_i$s. 
In Section (5), we will show that the link invariant and the 4--CB algebra follow for the
case of $G_2$.  
 

\mysec{$G_2$ IRF model.}

We wish to check our conjecture for the IRF Boltzmann weights, eqs. (1.20--1.21), for the $G_2$ IRF model.
The explicit Boltzmann weights of this model were given by Kuniba et al. \REF\Kuniba{A. Kuniba
and J. Suzuki, Phys.Lett A 160 (1991) 216.}\r\Kuniba. We wish to check our general ansatz, eqs. (1.20--1.21), specialized to the $G_2$ case.
The model is defined by taking for the RCFT $\cal O$ the WZW model based
on the Lie algebra $G_2$ at level $k$. For the field $h=v$ we substitute the fundamental representation
$[7]$, which is the $7$ dimensional representation. Thus the model is IRF$(G_2,[7],[7])$.
In the fusion product of $[h]$ and $[v]$, eq. (1.11), we encounter four representations,
$$[7]\times [7]=[1]+[14]+[27]+[7].\e$$
Thus, the $G_2$ theory is a four blocks theory. Note that we chose this order for the fields appearing in the
product, $\psi_0=[1]$, $\psi_1=[14]$, $\psi_2=[27]$ and $\psi_3=[7]$, to be  
consistent with the Boltzmann weights of Kuniba et al..

The dimensions of the fields in a WZW theory are given by
$$\Delta_\lambda={\lambda(\lambda+2 \rho)\over 2(k+g) },\e$$
where $\lambda$ is the highest weight of the representation, $\rho$ is half the sum of positive roots,
and $g$ is the dual Coxeter number.

We can now compute the crossing parameters $\zeta_i$, using eq. (1.19).  We find for the dimensions
of the fields $\psi_i$,
$$\Delta_0=0,\quad \Delta_1={12\over k+4},\quad \Delta_2={14\over k+4},\quad \Delta_3={6\over k+4},\e$$
where $\Delta_i$ is the dimension of the field $\psi_i$.

The crossing parameters are given, by  eq. (1.19),  $\zeta_i=\pi (\Delta_{i+1}-\Delta_i)/2$,
$$\zeta_0=\lambda=-{6\pi \over k+4},\quad \zeta_1=-{\pi\over k+4},\quad \zeta_2={4\pi\over k+4}.\e$$
Note that we inverted the signs of the crossing parameters. This is always allowed since in
the ansatz, eqs. (1.20--1.21), we can invert the signs of the crossing parameters along with the sign of $u$ and the resulting
Boltzmann weight is not changed (up to a possible sign). 

To calculate the Boltzmann weight $X_i(u)$ we need to know the braiding matrices of the RCFT
$\cal O$. Unfortunately, this has never been calculated directly. So our idea is to extract the 
braiding matrices from the solution of Kuniba et al., and then to compare that $X_i(u)$, as given
by eqs. (1.20--1.21) agrees with the Boltzmann weights of Kuniba et al. This will assure that our conjecture
is correct for this theory.

 We extract the projection operators $P_i^a$ from the above eqs. (3.4--3.6) and insert them into our ansatz
 of the Boltzmann weights, eqs. (1.20--1.21). Now we are in a position to compare our Boltzmann weights
 with those of Kuniba et al. We preform this calculation numerically by choosing $q=e^{i\pi/(k+4)}=0.7$.
 We compare all the Boltzmann weights for an arbitrary spectral parameter, $u$.
 We find a complete agreement. This illustrates that our ansatz is correct for $G_2$ theory.

{\bf 5.1 Kuperberg's $G_2$ link invariant.}

In ref. \r\Kuper,
Kuperberg introduced a tangle algebra for the $G_2$ link invariant. Our claim is that this
link invariant is identical, for the special case of $G_2$, to our link invariant in Section (4).
This proves the link invariant for this case, as well as the algebraic structure we find, namely
the 4--CB algebra, for this case.

Recall from eq. (5.4)  that the crossing multipliers for $G_2$ have the form
$$\zeta_0=-{6\pi\over k+4},\qquad \zeta_1=-{\pi\over k+4},\qquad \zeta_2={4\pi\over k+4}.$$
We find it convenient to define
$$q=e^{i\pi/(k+4)}.\e$$
Thus we find from eqs. (3.11--3.12) that the parameters of the algebra are given by
$$l=i q^{-12},\qquad b=-{[12][7][2]\over [6][4][1]},\e$$
where we defined,
$$[x]=q^x-q^{-x}.\e$$
To make contact with Kuperberg's work we also rescale $G_i^{\pm1}\rarrow i^{\pm1} G_i^{\pm1}$ and
$E_i\rarrow -E_i$. Also, his $q$ is our $q^2$.
We then find from eqs. (3.15--3.16),
$$G_i^2=-(-q^{14}+q^{12}-q^{10}+q^6-q^4+q^2) E_i+(q^{-2}+q^4-q^6)-(1-q^{-2}+q^6) G_i+q^4 G_i^{-1}.\e$$

Now, in Kalfagianni's work \r\Kal, the algebraic relations which follow from Kuperberg's tangle algebra were derived.
This author finds exactly the 4--CB algebra that we conjectured for the case of $G_2$,
with exactly the parameters $l,b,\alpha,\beta,\gamma,\delta$ that we calculated from our general ansatz.
This proves the 4--CB algebra for the case of $G_2$. It also shows the consistency of the link invariant,
described in Section (4), for this special case.

Actually in ref. \r\Kal\ additional relations are described. As will be shown in Sections  (6--7), these 
hold also for all 4--blocks theories.

\mysec{$H$ and $K$ relations}

In Kuperberg paper \r\Kuper, the diagrammatic operations $H$ and $K$ are defined.
Kalfagianni \r\Kal\ defined these algebraically.  There it was shown that in the
case of $G_2$ the operators $H$ and $K$ obey some far reaching algebraic relations. Our purpose here
is to generalize Kalfagianni's algebra to all the $4$-blocks lattice models. We will check these relations
with the $SU(2)$ fused 3x3 lattice models, assuming that if they hold both for $G_2$ and $SU(2)$, they
are correct generally. We prove these relations for any four blocks model in Section (7).

Our starting point are the relations, which hold for $G_2$,
$$ G_i-q^{-1} G_i^{-1}=(1-q^{-1})(H_i+(q+q^{-1}) E_i-1),\e$$
$$G_i-q G_i^{-1}=(1-q)(K_i-E_i+(q+q^{-1})).\e$$
where $q=\exp[i\pi /(2(k+4)]$, and $k$ is the level.
If we substitute the expressions fo $G_i$ and $G_i^{-1}$ in accordance with our ansatz eqs. (1.20--1.21),
we find that $H_i$ is proportional to $P_i^3$, the third projection operator. Our idea to generalize 
the operator $H_i$ by simply  equating it with this projection operator
$$H_i=P_i^3.\e$$

To express $H_i$ in terms of $G_i$, $G_i^{-1}$, $E_i$ and $1_i$ we simply solve the equations
for them, eqs. (3.4--3.6), using the relation $P_i^1=1_i-P_i^0-P_i^2-P_i^3$, to find $P_i^3$.
We thus get the relation,
\def\frac#1#2{{#1\over #2}}\def\text{}
$$\displaylines{z \, H_i =\left(
-  e^{i \zeta (0)+2 i \zeta (2)}+ e^{i \zeta 
(0)+2 i \zeta (1)+2 i \zeta (2)}+e^{i \zeta (0)+4 i \zeta 
(2)}-e^{i \zeta (0)+2 i \zeta (1)+4 i \zeta (2)}\right) E_i+
\cr
\left( -e^{2 i \zeta 
(1)+2 i \zeta (2)}-e^{2 i \zeta (0)+4 i \zeta (2)}+e^{2 i \zeta (0)+2 i 
\zeta (1)+4 i \zeta (2)}+e^{2 i \zeta (2)}\right) 1_i+
\cr
i e^{\frac{1}{2} i \zeta (0)+i 
\zeta (1)+3 i \zeta (2)} G_i-i e^{\frac{5}{2} i \zeta (0)+i \zeta (1)+5 i 
\zeta (2)} G_i+i e^{-\frac{1}{2} i \zeta (0)+i \zeta (1)+i \zeta (2)} 
\text{G_i^{-1}}-\cr \hfill
i e^{\frac{3}{2} i \zeta (0)+i \zeta (1)+3 i \zeta (2)} \text{G_i^{-1}}
,\hfill \enumber \cr}$$
where $z$ is defined by
$$z=\left(-1+e^{i \zeta (0)+i \zeta (2)}\right) \left(1+e^{i \zeta (0)+i \zeta 
(2)}\right) \left(-1+e^{2 i \zeta (1)+2 i \zeta (2)}\right) \left(1+e^{2 i 
\zeta (2)}\right)
.\e$$
It is convenient to define the parameters $r_j$ as the coefficients of $H_i$,
$$H_i=r_1 \,1_i+r_2\, E_i+r_3\, G_i+r_4\, G_i^{-1},\e$$
where $H_i$ is given by eqs. (6.4--6.5) above. Note that $H_i$ is defined for any values of the crossing
parameters $\zeta_i$ and thus for any four blocks theory.

We come now to the problem of defining $K_i$. We do this by imposing Kalfagianni's equation
which holds for $G_2$, 
$$E_{i\pm1} E_i H_{i\pm1}=E_{i\pm1} K_i,\e$$
and assuming that it holds for any theory and not just $G_2$. We substitute $H_i$ from eq. (6.4--6.5)
and use the relations of the BMW algebra eq. (3.18). We find that this relation holds if and
only if $K_i$ has the expression,
$$K_i=r_2\, 1_i+r_1\, E_i+r_4\, G_i+r_3\, G_i^{-1}.\e$$
Again, this relation is general for any four blocks theory.

We are now in a position to check Kalfagianni's  relations, which hold for $G_2$, to the other case
that we investigated which is the $SU(2)$ fused $3\times 3$ model. We do this using the explicit
Boltzmann weights which were given in ref. \r\Second. 

The crossing parameters for the $SU(2)$ $3\times 3$ model are given by \r\Second
$$\zeta_0={\pi\over k+2},\quad \zeta_1={2\pi\over k+2},\quad \zeta_2={3\pi\over k+2},\e$$
where $k$ is the level of the $SU(2)$ model.

We find that  all of Kalfagianni relations hold also for $SU(2)$ model. Below is a list of
the `simple' relations that hold not only for $G_2$ but also for $SU(2)$.

$$H_i E_i=0,\e$$
$$K_i E_i=E_i K_i=d  E_i,\e$$
$$K_i H_i=H_i K_i=c H_i,\e$$
$$K_i^2=a H_i+b K_i+e E_i+f,\e$$
$$H_i H_j=H_j H_i,\ {\rm if\ }|i-j|\geq 2,\e$$
$$K_i K_j=K_j K_i,\ {\rm if\ } |i-j|\geq 2,\e$$
$$G_{i\pm1} G_i H_{i\pm1}=H_i G_{i\pm1} G_i,\e$$
$$G_{i\pm1} G_i K_{i\pm1}=K_i G_{i\pm1} G_i,\e$$
$$E_i H_{i\pm1} E_i=x E_i,\e$$
$$E_i K_{i\pm1} E_i=0,\e$$
$$E_{i\pm1} E_i H_{i\pm1}=E_{i\pm1} K_i,\e$$
$$H_{i\pm1} E_i E_{i\pm1}=K_{i\pm1} E_i,\e$$
$$K_{i\pm1} E_i E_{i\pm1}=H_{i\pm1} E_i,\e$$
$$E_{i\pm1} E_i K_{i\pm1}=E_{i\pm1} H_i,\e$$
$$E_i H_{i\pm1} H_i=E_i K_{i\pm1} K_i,\e$$
$$H_{i\pm1} H_i E_{i\pm1}=K_{i\pm1} K_i E_{i\pm1},\e$$
$$K_i E_{i\pm1} H_i=H_{i\pm1} E_i K_{i\pm1},\e$$
$$H_i E_{i\pm1} H_i=K_{i\pm1} E_i K_{i\pm1}.\e$$

Some of these simple relations follow directly from the definition of $H_i$ and $K_i$ and the BMW
algebra, but not all. Here $a,b,c,d,e,f$ and $x$ are some coefficients which can be easily computed from the
definition eqs. (1.20--1.21). 

We get now to the 'complicated' relations, following Kalfagianni.  The relations are
$$\twoline{
H_i H_{i+1} H_i=H_{i+1} H_i H_{i+1}-v_1 (H_i E_{i+1} H_i-H_{i+1} E_i H_{i+1})-
}{
v_2(H_i K_{i+1} H_i-
H_{i+1} K_i H_{i+1})-v_3 (H_i-H_{i+1}).}$$

$$ \threeline{
H_i H_{i\pm1} K_i=K_{i\pm1} H_i H_{i\pm1}-x_1 (H_i K_{i\pm1} H_i-H_{i\pm1} K_i H_{i\pm1})-
}{
x_2(
H_i E_{i\pm1} H_i-H_{i\pm1} E_i H_{i\pm1})-x_3(H_i K_{i\pm1}-K_i H_{i\pm1})-
}{
x_4(K_{i\pm1} E_i-E_{i\pm1} K_i)-
x_5(H_i  E_{i\pm1}-E_i H_{i\pm1})-x_6(H_i-H_{i\pm1}).}$$

$$\threeline{
K_i K_{i+1} K_i=K_{i+1} K_i K_{i+1}-z_1 (H_i K_{i+1} H_i-H_{i+1} K_i H_{i+1})-
}{
z_2 (H_i K_{i+1}-
H_{i+1} K_i)-z_3 (K_{i+1} H_i-K_i H_{i+1})-z_4 (K_{i+1} E_i-K_i E_{i+1})-
}{
z_5 (E_i K_{i+1}-
E_{i+1} K_i)-z_6 (K_i-K_{i+1}).}$$

As was shown by Kalfagianni, these equations hold for the $G_2$ model for some values of the
parameters $v_i,x_i,z_i$. Our idea is to establish these relations for the $SU(2)$ $3\times 3$ model, for
some values of the parameters. We substituted the Boltzmann weights and solved for the 
parameters using some configurations. We found that these relations are obeyed also for $SU(2)$.
For the parameters we find the following general relations:
$$v_1=-r_2,\quad x_2=x_3=-r_1,\quad x_4=x_6, \e$$
where $r_1,r_2$ is given by eq. (6.6).
We also find that the coefficients $z_i$ are the same as the coefficients $x_i$,
$$z_1=x_1,\quad z_2=z_3=x_2=x_3,\quad z_4=z_5=x_4=x_6,\quad z_6=-x_5.\e$$

For example,  we give here the values of the parameters for $k+2=12$,
$$v_2=-\sqrt{2},\quad v_3=-\sqrt{1\over 12}.\e$$
$$x_1=\sqrt{3},\quad x_5={1\over 6}.\e$$
$$x_4=x_6=1/(\sqrt2 z)=\frac{\frac{1}{2}-\frac{i}{2}}{\sqrt{2} \left(-1+e^{\frac{i \pi }{3}}\right) \
\left(1+e^{\frac{i \pi }{3}}\right) \left(-1+e^{\frac{5 i \pi }{6}}\right)}.\e$$
$$x_2=x_3=-r_1=-(\sqrt 3/2+1)^{1/2}+1.\e$$

To get the parameters for general $k$ we solved these equations for general $q=\exp[\pi i/(k+2)]$ using
symbolic manipulation, substituting some configurations. We find the following expressions for $v_i$.
$$v_1= \frac{q^{11}+q^9+q^7}{q^{18}+q^{16}+2 q^{14}+3 
q^{12}+3 q^{10}+3 q^8+3 q^6+2 q^4+q^2+1},\e$$
$$v_2=
-\frac{q^{11}+q^9+q^7}{q^{18}+q^{16}+q^{14}+q^{12}+q^{10}+q^8+q^6+q^4+q^2+1},\e$$
$$v_3= -\frac{q^6 \
\left(q^{12}+q^{10}+q^8+q^6+q^4+q^2+1\right)}{\left(q^4+1\right) 
\left(q^6+1\right)^2 \left(q^8+q^6+q^4+q^2+1\right)}.\e$$

For $x_i$ we find the following expressions,
$$
x_1=\frac{q^{10}+q^6}{q^{16}+q^{12}+q^8+q^4+1},\e$$
$$x_2=x_3= -\frac{q^6}{q^{12}+q^8+q^6+q^4+1},\e$$
$$x_4=x_6= \frac{q^9 
\left(q^{12}+q^{10}+q^8+q^6+q^4+q^2+1\right)}{\left(q^2+1\right) 
\left(q^4+1\right) \left(q^{12}+q^8+q^6+q^4+1\right)^2},\e$$
$$
x_5=\frac{q^6 \
\left(q^{12}+q^{10}+q^8+q^6+q^4+q^2+1\right)^2}{\left(q^4-q^2+1\right)^2 
\left(q^6+q^4+q^2+1\right)^2 \left(q^8+q^6+q^4+q^2+1\right)^2}.\e$$

The $z_i$ are given by eq. (6.32) from $x_i$. We find that the relations eqs. (6.28--6.30) are indeed obeyed for any $q$.

Moreover, the following equations hold for the $SU(2)$ $3 \times 3$ model:
$$
\eqalign{
& K_iK_{i \pm 1}H_i= H_{i \pm 1} K_i K_{i \pm 1} - \frac{q^6  \left(q^4 - 1\right)  \left(q^{14} - 1\right)}{{\left(q^6 + 1\right)}^2  \left(q^8 - 1\right)  \left(q^{10} - 1\right)}(K_i E_{i \pm 1}-E_i K_{i \pm 1}) \cr
& - \frac{q^7  \left(q^4 - 1\right)  \left(q^6 - 1\right)}{\left(q^6 + 1\right)  \left(q^8 - 1\right)  \left(q^{10} - 1\right)}(K_{i \pm 1} H_i-H_{i \pm 1} K_i) \cr
& +\frac{q^7  \left(q^6 - 1\right)}{q^{20} - 1}(H_i K_{i \pm 1} H_i-H_{i \pm 1} K_i H_{i \pm 1}), } \e
$$
and
$$
\eqalign{
& K_{i + 1}H_iK_{i + 1}= K_i H_{i + 1} K_i + \cr
& + w_1 (H_i K_{i + 1}-K_i E_{i + 1} K_i-K_i H_{i + 1}+K_{i + 1} E_i K_{i + 1}+K_{i + 1} H_i-H_{i + 1} K_i) \cr
& + w_2 (K_{i + 1} E_i-E_{i + 1} K_i+H_i-H_{i + 1}+E_i K_{i + 1}-K_i E_{i + 1})\cr
& + w_3  (H_i K_{i + 1} H_i-H_{i + 1} K_i H_{i + 1}) + w_4 (E_i-E_{i + 1}), } \e
$$
where 
$$
\eqalign{
& w_1 = - \frac{\left(q^4 - 1\right)\left(q^{22} - 1\right)}{q\left(q^6 + 1\right)\left(q^8 - 1\right)\left(q^{10} - 1\right)}, \cr
& w_2 = \frac{\left(q^4 - 1\right)\left(q^4 - q^8\right)\left( - q^{36} + q^{22} + q^{14} - 1\right)}{q^2  {\left(q^6 + 1\right)}^2  {\left(q^8 - 1\right)}^2  {\left(q^{10} - 1\right)}^2}, \cr
& w_3 = \frac{q^{22} - 1}{q\left(q^{20} - 1\right)}, \cr
& w_4 = - \frac{{\left(q^4 - 1\right)}^2\left(q^{12} - 1\right)  {\left(q^{14} - 1\right)}^3\left(q^{22} - 1\right)}{q^3\left(q^6 - 1\right)  {\left(q^6 + 1\right)}^3  {\left(q^8 - 1\right)}^3  {\left(q^{10} - 1\right)}^3}.
}
$$

\mysec{$H$ and $K$ relations for the general four blocks models}

We write $s_i = e^{\sqrt{-1} \zeta_{i}}$, $i \in \{0,1,2,3\}$. Let 
$$\eqalign{
G_i & = 2^3 s_0^{-\frac{3}{2}} \sin(\zeta_0) \sin(\zeta_1) \sin(\zeta_2) X_i \cr
& = \frac{ \sqrt{-1}P_{i}^{1}}{\sqrt{s_0} s_1 s_2} - \frac{ \sqrt{-1}P_{i}^{0}}{{s_0}^{\frac{5}{2}} s_1 s_2} + \frac{ \sqrt{-1}s_1 s_2 P_{i}^{3}}{\sqrt{s_0}} - \frac{ \sqrt{-1}s_1 P_{i}^{2} }{\sqrt{s_0} s_2}. }
$$

Then 
$$\eqalign{
G_i^{-1} & = - \frac{ \sqrt{-1}\sqrt{s_0} P_{i}^{3} }{s_1 s_2} -  \sqrt{-1}\sqrt{s_0} s_1 s_2 P_{i}^{1}  + \sqrt{-1} {s_0}^{\frac{5}{2}} s_1 s_2 P_{i}^{0}  + \frac{ \sqrt{-1}\sqrt{s_0} s_2 P_{i}^{2} }{s_1}, \cr
G_i^2 & = - \frac{P_{i}^{1}}{s_0 {s_1}^2 {s_2}^2} - \frac{{s_1}^2 P_{i}^{2}}{s_0 {s_2}^2} - \frac{{s_1}^2 {s_2}^2 P_{i}^{3}}{s_0} - \frac{P_{i}^{0}}{{s_0}^5 {s_1}^2 {s_2}^2}. }
$$

Let 
$$
E_i = X_i(\zeta_0) = \frac{\left({s_0}^2 {s_1}^2 - 1\right) \left({s_0}^2 {s_2}^2 - 1\right) \left({s_0}^2 + 1\right) }{\left({s_1}^2 - 1\right) \left({s_2}^2 - 1\right) {s_0}^3} P_{i}^0.
$$

The skein relation is $\sum_a P_i^a = 1$. That is,
$$\eqalign{
G_i^2 &  =\frac{E_i \left({s_0}^2 {s_1}^2 {s_2}^2 + 1\right) \left({s_1}^2 - 1\right) \left({s_2}^2 - 1\right)}{\left({s_0}^2 {s_2}^2 - 1\right) {s_0}^2 {s_1}^2 {s_2}^2}  \cr 
& + \frac{ \sqrt{-1}G_{i} \left({s_1}^2 {s_2}^2  - {s_1}^2 + 1\right)}{\sqrt{s_0} s_1 s_2} - \frac{\left({s_1}^2 {s_2}^2 - {s_2}^2 + 1\right)}{s_0 {s_2}^2} + \frac{ \sqrt{-1}s_1 G^{-1}_{i} }{{s_0}^{\frac{3}{2}} s_2}. }
$$

We also have 
$$\eqalign{
G_i^{-2} & = - \frac{s_0 \left({s_1}^2 {s_2}^2 - {s_1}^2 + 1\right)}{{s_1}^2}- \frac{\left({s_0}^2 {s_1}^2 {s_2}^2 + 1\right) \left({s_1}^2 - 1\right) \left({s_2}^2 - 1\right) {s_0}^2}{\left({s_0}^2 {s_2}^2 - 1\right) {s_1}^2}E_i \cr
& - \frac{ \sqrt{-1}\sqrt{s_0} \left({s_1}^2 {s_2}^2  - {s_2}^2  + 1\right)}{s_1 s_2}G^{-1}_{i}- \frac{ \sqrt{-1}{s_0}^{\frac{3}{2}} s_2 }{s_1}G_{i}. }
$$

Let $H_i = P_i^3 =r_1 1_i + r_2 E_i + r_3 G_i + r_4 G_i^{-1}$ and $K_i = r_2 E_i + r_1 1_i + r_4 G_i + r_3 G_i^{-1}$. Then
$$\eqalign{
& r_1 = \frac{\left({s_1}^2 - 1\right) {s_2}^2}{\left({s_1}^2 {s_2}^2 - 1\right) \left({s_2}^2 + 1\right)}, \cr
& r_2 = - \frac{\left({s_1}^2 - 1\right) \left({s_2}^2 - 1\right) s_0 {s_2}^2}{\left({s_0}^2 {s_2}^2 - 1\right) \left({s_1}^2 {s_2}^2 - 1\right) \left({s_2}^2 + 1\right)}, \cr
& r_3 = - \frac{ \sqrt{-1}\sqrt{s_0} s_1 {s_2}^3 }{\left({s_1}^2 {s_2}^2 - 1\right) \left({s_2}^2 + 1\right)}, \cr
& r_4 = - \frac{ \sqrt{-1}s_1 s_2 }{\left({s_1}^2 {s_2}^2 - 1\right) \left({s_2}^2 + 1\right) \sqrt{s_0}}.  }
$$

We have that
$$\eqalign{
G_i  & = \frac{E_i \left(r_1 r_4 - r_2 r_3\right)}{\left({r_3}^2 - {r_4}^2\right)} - \frac{\left(r_1 r_3 - r_2 r_4\right)}{\left({r_3}^2 - {r_4}^2\right)} + \frac{H_i r_3}{\left({r_3}^2 - {r_4}^2\right)} - \frac{K_i r_4}{\left({r_3}^2 - {r_4}^2\right)} \cr
& = - \frac{ \sqrt{-1}{s_0}^{\frac{3}{2}} s_2 \left({s_1}^2  - 1\right)}{\left({s_0}^2 {s_2}^2 - 1\right) s_1} + \frac{ \sqrt{-1}E_i \sqrt{s_0} s_2 \left({s_1}^2  - 1\right)}{\left({s_0}^2 {s_2}^2 - 1\right) s_1} \cr
& \quad + \frac{ \sqrt{-1}H_i \left({s_1}^2 {s_2}^2  - 1\right) \left({s_2}^2 + 1\right) {s_0}^{\frac{3}{2}} s_2}{\left({s_0}^2 {s_2}^4 - 1\right) s_1} - \frac{ \sqrt{-1}K_i \left({s_1}^2 {s_2}^2  - 1\right) \left({s_2}^2 + 1\right) \sqrt{s_0}}{\left({s_0}^2 {s_2}^4 - 1\right) s_1 s_2}, }
$$
$$\eqalign{
G_i^{-1} & =
\frac{\left(r_1 r_4 - r_2 r_3\right)}{\left({r_3}^2 - {r_4}^2\right)} - \frac{E_i \left(r_1 r_3 - r_2 r_4\right)}{\left({r_3}^2 - {r_4}^2\right)} - \frac{H_i r_4}{\left({r_3}^2 - {r_4}^2\right)} + \frac{K_i r_3}{\left({r_3}^2 - {r_4}^2\right)} \cr
& =  \frac{ \sqrt{-1}\sqrt{s_0} s_2 \left({s_1}^2  - 1\right)}{\left({s_0}^2 {s_2}^2 - 1\right) s_1} - \frac{ \sqrt{-1}E_i {s_0}^{\frac{3}{2}} s_2 \left({s_1}^2 - 1\right)}{\left({s_0}^2 {s_2}^2 - 1\right) s_1} \cr
& \quad  + \frac{ \sqrt{-1}K_i \left({s_1}^2 {s_2}^2  - 1\right) \left({s_2}^2 + 1\right) {s_0}^{\frac{3}{2}} s_2}{\left({s_0}^2 {s_2}^4 - 1\right) s_1} - \frac{ \sqrt{-1}H_i \left({s_1}^2 {s_2}^2  - 1\right) \left({s_2}^2 + 1\right) \sqrt{s_0}}{\left({s_0}^2 {s_2}^4 - 1\right) s_1 s_2}.  }
$$

We have $X_i(u) = p_1 A + p_2 A^{-1} + p_3 A^{-3} + p_4 A^3$, where $A=e^{iu}$ and
$$\eqalign{
&  p_1 =  \frac{E_i \left({s_1}^2 - 1\right) s_0 \left({s_0}^2 {s_2}^4 + {s_2}^2 + 1\right)}{\left({s_0}^2 {s_2}^2 - 1\right)} - \frac{\left({s_1}^2 - 1\right) \left({s_0}^2 {s_2}^4 + {s_0}^2 {s_2}^2 + 1\right)}{\left({s_0}^2 {s_2}^2 - 1\right)} \cr
& \quad + \frac{H_i \left({s_1}^2 {s_2}^2 - 1\right) \left({s_2}^2 + 1\right) \left({s_0}^2 {s_2}^2 + {s_0}^2 + 1\right)}{\left({s_0}^2 {s_2}^4 - 1\right)}  \cr
& \quad - \frac{K_i \left({s_1}^2 {s_2}^2 - 1\right) \left({s_2}^2 + 1\right) s_0 \left({s_0}^2 {s_2}^2 + {s_2}^2 + 1\right)}{\left({s_0}^2 {s_2}^4 - 1\right)},  }
$$
$$\eqalign{
& p_2 =
\frac{\left({s_1}^2 - 1\right) {s_0}^2 \left({s_0}^2 {s_2}^4 + {s_2}^2 + 1\right)}{\left({s_0}^2 {s_2}^2 - 1\right)} - \frac{E_i \left({s_1}^2 - 1\right) s_0 \left({s_0}^2 {s_2}^4 + {s_0}^2 {s_2}^2 + 1\right)}{\left({s_0}^2 {s_2}^2 - 1\right)} \cr
& \quad - \frac{H_i \left({s_1}^2 {s_2}^2 - 1\right) \left({s_2}^2 + 1\right) {s_0}^2 \left({s_0}^2 {s_2}^2 + {s_2}^2 + 1\right)}{\left({s_0}^2 {s_2}^4 - 1\right)} \cr
& \quad + \frac{K_i \left({s_1}^2 {s_2}^2 - 1\right) \left({s_2}^2 + 1\right) s_0 \left({s_0}^2 {s_2}^2 + {s_0}^2 + 1\right)}{\left({s_0}^2 {s_2}^4 - 1\right)}, }
$$
$$\eqalign{
&   p_3 =
\frac{E_i \left({s_1}^2 - 1\right) {s_0}^3 {s_2}^2}{\left({s_0}^2 {s_2}^2 - 1\right)} - \frac{K_i \left({s_1}^2 {s_2}^2 - 1\right) \left({s_2}^2 + 1\right) {s_0}^3}{\left({s_0}^2 {s_2}^4 - 1\right)} \cr
& \quad - \frac{\left({s_1}^2 - 1\right) {s_0}^4 {s_2}^2}{\left({s_0}^2 {s_2}^2 - 1\right)} + \frac{H_i \left({s_1}^2 {s_2}^2 - 1\right) \left({s_2}^2 + 1\right) {s_0}^4 {s_2}^2}{\left({s_0}^2 {s_2}^4 - 1\right)} \cr
& p_4 = 
\frac{\left({s_1}^2 - 1\right) {s_2}^2}{\left({s_0}^2 {s_2}^2 - 1\right)} - \frac{H_i \left({s_1}^2 {s_2}^2 - 1\right) \left({s_2}^2 + 1\right)}{\left({s_0}^2 {s_2}^4 - 1\right)} \cr
& \quad - \frac{E_i \left({s_1}^2 - 1\right) s_0 {s_2}^2}{\left({s_0}^2 {s_2}^2 - 1\right)} + \frac{K_i \left({s_1}^2 {s_2}^2 - 1\right) \left({s_2}^2 + 1\right) s_0 {s_2}^2}{\left({s_0}^2 {s_2}^4 - 1\right)}. }
$$

Assume that $G_i, G_i^{-1}, E_i, 1_i$ satisfy the following relations (which are BMW and the skein relations):

$$
\eqalign{
& G_i G_j = G_j G_i, \quad E_i G_j = G_j E_i, \quad |i-j| \ge 2, \cr
& G_i G_j G_i = G_j G_i G_j, \quad |i-j|=1, \cr
& E_i E_{i \pm 1} E_i = E_i, \cr
& G_i E_i = E_iG_i = - \frac{   \sqrt{-1}}{{s_0}^{\frac{5}{2}}  s_1  s_2} E_i, \cr
& E_i G_j E_i =   \sqrt{-1} {s_0}^{\frac{5}{2}}  s_1  s_2   E_i, \quad |i-j|=1, \cr
& E_i^2 = \frac{  \left({s_0}^2  {s_1}^2 - 1\right)  \left({s_0}^2  {s_2}^2 - 1\right)  \left({s_0}^2 + 1\right)}{\left({s_1}^2 - 1\right)  \left({s_2}^2 - 1\right)  {s_0}^3}E_i \cr
& E_i G_{i \pm 1} G_i = G_{i \pm 1} G_i E_{i \pm 1} = E_i E_{i \pm 1}, \cr
& G_{i \pm 1} E_i E_{i \pm 1} = G_i^{-1} E_{i \pm 1}, \cr
& E_i E_{i \pm 1} G_i = E_i G_{i \pm 1}^{-1}, \cr
& G_{i \pm 1}^{-1} E_i G_{i \pm 1}^{-1} = G_i E_{i \pm 1} G_i, }
$$
and 
$$
\eqalign{
G_i^2 &=\frac{E_i \left({s_0}^2 {s_1}^2 {s_2}^2 + 1\right) \left({s_1}^2 - 1\right) \left({s_2}^2 - 1\right)}{\left({s_0}^2 {s_2}^2 - 1\right) {s_0}^2 {s_1}^2 {s_2}^2} + \frac{ \sqrt{-1}G_{i} \left({s_1}^2 {s_2}^2 - {s_1}^2  + 1\right)}{\sqrt{s_0} s_1 s_2} \cr
&- \frac{\left({s_1}^2 {s_2}^2 - {s_2}^2 + 1\right)}{s_0 {s_2}^2} 1_i + \frac{ \sqrt{-1}s_1 G^{-1}_{i} }{{s_0}^{\frac{3}{2}} s_2}, }
$$
$$\eqalign{
G_i^{-2} & = - \frac{s_0 \left({s_1}^2 {s_2}^2 - {s_1}^2 + 1\right)}{{s_1}^2} 1_i - \frac{\left({s_0}^2 {s_1}^2 {s_2}^2 + 1\right) \left({s_1}^2 - 1\right) \left({s_2}^2 - 1\right) {s_0}^2}{\left({s_0}^2 {s_2}^2 - 1\right) {s_1}^2}E_i \cr
& - \frac{ \sqrt{-1}\sqrt{s_0} \left({s_1}^2 {s_2}^2  - {s_2}^2  + 1\right)}{s_1 s_2}G^{-1}_{i}- \frac{ \sqrt{-1}{s_0}^{\frac{3}{2}} s_2 }{s_1}G_{i}. }
$$

It is easy to check the following.
$$\eqalign{
& H_iE_i=E_iH_i=0,   \cr
& K_iE_i=E_iK_i= q_1 E_i, \cr
& E_i^2 = \frac{ \left({s_0}^2  {s_1}^2 - 1\right)  \left({s_0}^2  {s_2}^2 - 1\right)  \left({s_0}^2 + 1\right)}{\left({s_1}^2 - 1\right)  \left({s_2}^2 - 1\right)  {s_0}^3} E_i, \cr
& K_iH_i=H_iK_i=q_2H_i, } \e
$$
$$\eqalign{
& H_i^2=H_i, \cr
& K_i^2 = v_1 K_i + v_2 H_i + v_3 1_i + v_4 E_i, \cr
& E_i K_{i \pm 1} E_i = 0, \cr
& E_i E_{i \pm 1} H_i = E_i K_{i \pm 1}, \cr
& E_i E_{i \pm 1} K_i = E_i H_{i \pm 1}, \cr
& H_i E_{i \pm 1} E_i = K_{i \pm 1} E_i, \cr
& K_i E_{i \pm 1} E_i = H_{i \pm 1} E_i, \cr
& E_i H_{i \pm 1} E_i = q_1 E_{i}, } \e
$$
$$\eqalign{
& E_i H_{i \pm 1} H_i = q_2 E_i K_{i \pm 1} = E_i K_{i \pm 1} K_i, \cr
& H_i H_{i \pm 1} E_i =  q_2 K_{i \pm 1} E_i = K_i K_{i \pm 1} E_i, \cr
& K_i E_{i \pm 1} H_i = H_{i \pm 1} E_i K_{i \pm 1}, \cr
& H_i E_{i \pm 1} H_i = K_{i \pm 1} E_i K_{i \pm 1}, \cr
& E_i H_{i+1} K_i - E_{i+1} H_i K_{i+1} = E_i(v_1 H_{i+1} + v_2 K_{i+1} + v_3 E_{i+1} + v_4 1_i) \cr
& \qquad \qquad \qquad \qquad \qquad \qquad \qquad \qquad + E_{i+1}( - v_1 H_i - v_2 K_i - v_3 E_i - v_4 1_i),
\cr 
& K_i H_{i+1} E_i - K_{i+1} H_i E_{i+1} = (v_1 H_{i+1} + v_2 K_{i+1} + v_3 E_{i+1} + v_4 1_i)E_i  \cr
& \qquad \qquad \qquad \qquad \qquad \qquad \qquad \qquad + ( - v_1 H_i - v_2 K_i - v_3 E_i - v_4 1_i)E_{i+1}, \cr
& H_i K_{i \pm 1} E_i - E_{i \pm 1} K_i H_{i \pm 1} = K_{i \pm 1} E_i - E_{i \pm 1} K_i, \cr
& E_iH_{i \pm 1}K_i - K_i H_{i \pm 1} E_i = E_i( v_1 H_i + v_2 K_i + v_3 E_i ) - (v_1 H_i + v_2 K_i + v_3 E_i) E_i. } \e
$$
where
$$\eqalign{
& q_1 = - \frac{ \left({s_0}^2 {s_1}^2 {s_2}^2 - 1\right) \left( - {s_0}^6 {s_2}^6 + {s_0}^4 {s_2}^2 + {s_0}^2 {s_2}^4 - 1\right)}{\left({s_0}^2 {s_2}^2 - 1\right) \left({s_1}^2 {s_2}^2 - 1\right) \left({s_2}^4 - 1\right) {s_0}^3}, \cr
& q_2 = \frac{\left({s_1}^2 - {s_0}^2 {s_2}^2\right) \left({s_0}^2 - 1\right) {s_2}^2}{\left({s_0}^2 {s_2}^2 - 1\right) \left({s_1}^2 {s_2}^2 - 1\right) \left({s_2}^2 + 1\right) s_0}, }
$$
$$\eqalign{
& v_1 = \frac{\left({s_1}^2 - 1\right) \left( - {s_0}^4 {s_2}^6 - {s_0}^2 {s_2}^4 + {s_0}^2 {s_2}^2 + 1\right)}{\left({s_0}^2 {s_2}^2 - 1\right) \left({s_1}^2 {s_2}^2 - 1\right) \left({s_2}^2 + 1\right) s_0},\cr 
& v_2 = \frac{\left({s_0}^2 {s_2}^2 + 1\right) \left({s_1}^2 - {s_0}^2 {s_2}^2\right)}{\left({s_1}^2 {s_2}^2 - 1\right) \left({s_2}^2 + 1\right) {s_0}^2},\cr 
& v_3 = \frac{{\left({s_0}^2 {s_2}^4 - 1\right)}^2 \left({s_0}^4 {s_1}^2 {s_2}^4 - {s_0}^2 {s_1}^4 {s_2}^2 - {s_0}^2 {s_2}^2 + {s_1}^2\right)}{{\left({s_0}^2 {s_2}^2 - 1\right)}^2 {\left({s_1}^2 {s_2}^2 - 1\right)}^2 {\left({s_2}^2 + 1\right)}^2 {s_0}^2},\cr 
& v_4 = \frac{\left({s_0}^2 {s_2}^2 + 1\right) {\left({s_0}^2 {s_2}^4 - 1\right)}^2 \left({s_0}^2 {s_1}^2 {s_2}^2 - 1\right) \left({s_1}^2-1\right)}{\left({s_0}^2 {s_2}^2 - 1\right) {\left({s_1}^2 {s_2}^2 - 1\right)}^2 \left({s_2}^2 - 1\right) {\left({s_2}^2 + 1\right)}^2 {s_0}^3}. }
$$

Assume that $H_i, K_i, E_i$ satisfy (7.1), (7.2), (7.3). 
Then the Yang-Baxter equation $$X_i(u)X_{i+1}(u+v)X_i(v) = X_{i+1}(v)X_{i}(u+v)X_{i+1}(u)$$ is equivalent to the following equations.

$$\eqalign{
H_iH_{i+1}H_i = & H_{i+1}H_iH_{i+1} + a_1 (H_i-H_{i+1}) \cr
& + a_2(K_iE_{i+1}K_i - K_{i+1}E_iK_{i+1}) + a_3(H_iK_{i+1}H_i-H_{i+1}K_iH_{i+1}),
}  \e
$$
$$\eqalign{
K_iK_{i \pm 1}H_i= &H_{i \pm 1}K_iK_{i \pm 1} + a_1(E_i K_{i \pm 1} - K_i E_{i \pm 1}) \cr
& + a_2 (H_{i \pm 1}K_i-K_{i \pm 1}H_i) + a_3(H_i K_{i \pm 1} H_i-H_{i \pm 1}K_iH_{i \pm 1}),
}  \e
$$
$$\eqalign{
K_{i+1}K_iK_{i+1} & = K_iK_{i +1}K_i+b_1 (E_iK_{i +1}-K_iE_{i +1}+K_{i +1}E_i-E_{i +1}K_i) \cr
& + b_2 (-H_{i +1}K_i+K_{i +1}H_i+H_iK_{i +1}-K_iH_{i +1}) \cr
&+ b_3(H_iK_{i +1}H_i-H_{i +1}K_iH_{i +1})+b_4(-K_i+K_{i +1}), 
}  \e
$$
$$\eqalign{
K_iH_{i \pm 1}H_i&=  H_{i \pm 1}H_iK_{i \pm 1}  \cr
&  + b_1(K_iE_{i \pm 1}-E_iK_{i \pm 1}-H_i+H_{i \pm 1})  \cr
& + b_2(-K_{i \pm 1}H_i+H_{i \pm 1}K_i+K_iE_{i \pm 1}K_i-K_{i \pm 1}E_iK_{i \pm 1}) \cr
& + b_3( -H_iK_{i \pm 1}H_i+H_{i \pm 1}K_iH_{i \pm 1}) \cr
& b_4(H_{i \pm 1}E_i-E_{i \pm 1}H_i),
}  \e
$$
$$\eqalign{
K_{i+1}H_iK_{i+1}&= K_iH_{i +1}K_i \cr
& + c_1(-H_iK_{i +1}+K_iE_{i +1}K_i+K_iH_{i +1}-K_{i +1}E_iK_{i +1}-K_{i +1}H_i+H_{i +1}K_i) \cr
& +c_2(-K_{i +1}E_i+E_{i +1}K_i-H_i+H_{i +1}-E_iK_{i +1}+K_iE_{i +1}) \cr
& + c_3 (H_iK_{i +1}H_i-H_{i +1}K_iH_{i +1}
)+c_4(-E_i+E_{i +1}),
}  \e
$$
where 
$$\eqalign{
& a_1 = \frac{({s_0}^2 {s_2}^4 - 1) ({s_1}^2 - 1) {s_2}^2}{({s_0}^2 {s_2}^2 - 1) ({s_1}^2 {s_2}^2 - 1) {({s_2}^2 +1)}^2}, \cr
& a_2 = \frac{({s_1}^2 - 1) ({s_2}^2 - 1) s_0 {s_2}^2}{({s_0}^2 {s_2}^2 - 1) ({s_1}^2 {s_2}^2 - 1) ({s_2}^2 +1)}, \cr
& a_3 = \frac{({s_2}^2 - 1) s_0 {s_2}^2}{{s_0}^2 {s_2}^6 - 1},
}  
$$
$$\eqalign{
& b_1 = - \frac{({s_0}^2 {s_2}^4 - 1) ({s_1}^2 - {s_0}^2 {s_2}^2) ({s_1}^2 - 1) {s_2}^2}{({s_0}^2 {s_2}^2 - 1) {({s_1}^2 {s_2}^2 - 1)}^2 {({s_2}^2 +1)}^2 s_0},  \cr
& b_2 = - \frac{({s_1}^2 - 1) {s_2}^2}{({s_1}^2 {s_2}^2 - 1) ({s_2}^2 +1)},  \cr
& b_3 = - \frac{({s_2}^2 - {s_0}^2 {s_2}^4)}{({s_0}^2 {s_2}^6 - 1)},  \cr
& b_4 = \frac{{({s_0}^2 {s_2}^4 - 1)}^2 {({s_1}^2 - 1)}^2 {s_2}^2}{{({s_0}^2 {s_2}^2 - 1)}^2 {({s_1}^2 {s_2}^2 - 1)}^2 {({s_2}^2 +1)}^2},
} 
$$
$$\eqalign{
& c_1 =\frac{({s_0}^4 {s_2}^6 - 1) ({s_1}^2 - 1)}{({s_0}^2 {s_2}^2 - 1) ({s_1}^2 {s_2}^2 - 1) ({s_2}^2 +1) s_0}, \cr
& c_2 =- \frac{({s_1}^2 - {s_0}^2 {s_2}^2) ({s_1}^2 - 1) ( - {s_0}^6 {s_2}^{10} + {s_0}^4 {s_2}^6 + {s_0}^2 {s_2}^4 - 1)}{{({s_0}^2 {s_2}^2 - 1)}^2 {({s_1}^2 {s_2}^2 - 1)}^2 {({s_2}^2 +1)}^2 {s_0}^2},   \cr
& c_3 =\frac{{s_0}^4 {s_2}^6 - 1}{{s_0}^3 {s_2}^6 - s_0},  \cr
& c_4 = \frac{{({s_0}^2 {s_2}^4 - 1)}^3 ({s_0}^4 {s_2}^6 - 1) ({s_0}^2 {s_1}^2 {s_2}^2 - 1) {({s_1}^2 - 1)}^2}{{({s_0}^2 {s_2}^2 - 1)}^3 {({s_1}^2 {s_2}^2 - 1)}^3 ({s_2}^2 - 1) {({s_2}^2 +1)}^3 {s_0}^3} .
} 
$$

Let ${\fam\bbb K}$ be a field. We define a ${\fam\bbb K}(s_0, s_1, s_2)$-algebra $A$ generated by $H_i, K_i, E_i$, $i \in \{1, \ldots, n\}$ subject to the relations (7.1--7.8). We have the following theorem. 

Suppose that $H_i, K_i, E_i$, $i \in \{1, \ldots, n\}$ satisfy (7.1--7.3). Then $X_i(u)$ satisfies YBE if and only if $H_i, K_i, E_i$ satisfy the other defining relations in the algebra $A$.

When $s_0=q$, $s_1=q^2$, $s_2=q^3$, the equations (7.4--7.8) are (6.28), (6.44), (6.30), (6.29), and (6.45) respectively.

Since we showed in Section (3) that the YBE and BMW are obeyed if and only if the relation $g$, eqs. (3.19--3.21), is obeyed, it follows that this relation is equivalent to the algebra $A$ or to the general $H$ and $K$ relations.

\mysec{Conclusions.}

We studied in this paper the algebraic structure underlining solvable IRF (Interaction Round the Face)
lattice models. We proposed that the algebra depends only on the number of blocks.
For two blocks we obtain the Hecke $A_N$ algebra and the Temperley--Lieb (TL) algebra. For three blocks, we found that it is the weak
Birman--Murakami--Wenzl (BMW) algebra. This algebra has a quotient which is the BMW algebra, which contains the TL algebra, with a different skein
relation. In fact, in all cases we checked the all the BMW relations hold. We conjecture that this phenomenon is general.
For four blocks, we found that the algebra is a version of BMW algebra, with a different skein relation, along with
one additional relation. This algebra is new, to the best of our knowledge, and we plan to further investigate it.

These results suggest the following general picture.
The $n$ block algebras, for $n=1,2,...$ form a chain of algebras, which are quotients of the universal free algebra with generators $1,E_i,G_i,$ such that the relations which define the $n+1$th quotient contain those of the $n$th quotient, except for the skein relation which is different. We find that this picture is established for the two, three and four blocks cases.

To further investigate this chain of algebras, it behooves us to study the five blocks algebra, and hopefully
also higher blocks, with the ultimate goal of finding the general algebra. This we intend to pursue in future work.

If the conjecture of Section (4) holds, then our results will be relevant to knot theory, since then the algebras can be used to define new link invariants.
In addition, this algebraic structure  we found sheds more light into the physics of IRF models.

\ack
We thank Hans Wenzl for very helpful discussions. D.G. thank the theory department of CERN for
the hospitality while part of this work was done. We thank Ida Deichaite for remarks on the manuscript.
R.T. was supported by a research grant from the center for new scientists, Weizmann Institute of Science. J.L. was supported by the Austrian Science Fund (FWF): M 2633-N32 Meitner Program.


\refout


\def\frac#1#2{{#1\over #2}}
\APPENDIX{A}{A}

\def\to{=}
\def\text#1{#1}
\overfullrule=0pt
\line{\bf Weights of the $G_2$ model.\hfill}

The Boltzmann weights are taken from Kuniba and Suzuki, ref. \r\Kuniba.

The Weyl vector $\rho =(1,1)$ and the weights of the seven-dimensional representation space of $U_q(G_2)$ are
$$\manyeq{
&e_{-3}=(0,1)\,,\; e_{-2}=(1,-1)\,,\cr
&e_{-1}=(-1,2)\,,\; e_{0}=(0,0)\,,
}$$
and 
$$
e_{i}=-e_{-i} \,.
$$

We consider the highest weight module of $U_q(G_2)$ with the highest weight
$$\manyeq{
&a=(a_1,a_2)\,. \cr
}$$
We introduce function 
$$\manyeq{
&G[a]=s[3a_1]s[a_2]s[3a_1+a_2]s[3a_1+2a_2]s[3a_1+3a_2]s[6a_1+3a_2]\,.
}$$
Now we introduce the Boltzmann weights. First of all, for $\mu\neq 0$ 
$$\manyeq{
&B[a,u,\mu,\mu,\mu,\mu]=\frac{s[1+u]s[4+u]s[6+u]}{s[1]s[4]s[6]}
}$$
$$\manyeq{
&B[a,u,-3,-2,-3,-2]=\frac{s[a_2-u]s[4+u]s[6+u]}{s[a_2]s[4]s[6]}\cr
&B[a,u,-2,-3,-3,-2]=\left(\frac{s[a_2+1]s[a_2-1]}{s[a_2]^2}\right)^{\frac{1}{2}}\frac{s[u]s[4+u]s[6+u]}{s[1]s[4]s[6]}\cr
&B[a,u,-3,-1,-3,-1]=\frac{s[3a_1+a_2-u]s[4+u]s[6+u]}{s[3a_1+a_2]s[4]s[6]}\cr
&B[a,u,-1,-3,-3,-1]=\left(\frac{s[3a_1+a_2+1]s[3a_1+a_2-1]}{s[3a_1+a_2]^2}\right)^{\frac{1}{2}}\frac{s[u]s[4+u]s[6+u]}{s[1]s[4]s[6]}\cr
&B[a,u,-2,1,-2,1]=\frac{s[3a_1+2a_2-u]s[4+u]s[6+u]}{s[3a_1+2a_2]s[4]s[6]}\cr
&B[a,u,1,-2,-2,1]=\left(\frac{s[3a_1+2a_2+1]s[3a_1+2a_2-1]}{s[3a_1+2a_2]^2}\right)^{\frac{1}{2}}\frac{s[u]s[4+u]s[6+u]}{s[1]s[4]s[6]}\cr
&B[a,u,-3,0,-3,0]=\frac{s[3a_1+a_2-2-u]s[a_2-u]s[6+u]}{s[3a_1+a_2-2]s[a_2]s[6]}+\cr
&\frac{s[a_2+3]s[3a_1+a_2-1]s[3a_1+a_2+2]s[3a_1+2a_2-2-u]s[2]s[u]s[6+u]}{s[a_2]s[3a_1+a_2]s[3a_1+a_2-2]s[3a_1+2a_2+1]s[1]s[4]s[6]}\cr
&B[a,u,0,-3,-3,0]=\frac{s[u]s[3+u]s[6+u]}{s[1]s[4]s[6]}\times\cr
&\times\left(\frac{s[a_2-1]s[a_2+2]s[3a_1+a_2-1]s[3a_1+a_2+2]s[3a_1+2a_2-1]s[3a_1+2a_2+3]}{s[a_2]s[a_2+1]s[3a_1+a_2]s[3a_1+a_2+1]s[3a_1+2a_2+1]^2}\right)^{\frac{1}{2}}
\cr
&B[a,u,-2,-1,0,-3]=\frac{s[a_2+3+u]s[u]s[6+u]}{s[3a_1+a_2+1]s[4]s[6]}\times\cr
&\times\left(\frac{s[3a_1+3]s[3a_1+a_2-1]s[3a_1+a_2+2]s[3a_1+2a_2-1]s[2]}{s[3a_1]s[3a_1+2a_2+1]s[a_2]s[a_2+1]s[1]}\right)^{\frac{1}{2}}
\cr
&B[a,u,-1,-2,0,-3]=-\frac{s[3a_1+a_2+3+u]s[u]s[6+u]}{s[a_2+1]s[4]s[6]}\times\cr
&\times\left(\frac{s[3a_1-3]s[a_2-1]s[a_2+2]s[3a_1+2a_2-1]s[2]}{s[3a_1]s[3a_1+2a_2+1]s[3a_1+a_2]s[3a_1+a_2+1]s[1]}\right)^{\frac{1}{2}}
}$$
$$\manyeq{
&B[a,u,-2,-1,-2,-1]=\cr
&\frac{s[3a_1-u]s[2+u]s[6+u]}{s[3a_1]s[2]s[6]}+\frac{s[a_2-2]s[3a_1+a_2+3]s[3a_1-2-u]s[u]s[6+u]}{s[a_2]s[3a_1+a_2+1]s[3a_1]s[4]s[6]}\cr
&B[a,u,-3,1,-3,1]=\frac{s[3a_1+3a_2-u]s[2+u]s[6+u]}{s[3a_1+3a_2]s[2]s[6]}+\cr
&\frac{s[a_2+2]s[3a_1+2a_2+3]s[3a_1+3a_2-2-u]s[u]s[6+u]}{s[a_2]s[3a_1+2a_2+1]s[3a_1+3a_2]s[4]s[6]}\cr
&B[a,u,1,-3,0,-2]=-\frac{s[3a_1+2a_2+3+u]s[u]s[6+u]}{s[a_2-1]s[4]s[6]}\times\cr
&\times\left(\frac{s[a_2-2]s[a_2+1]s[3a_1+a_2-1]s[3a_1+3a_2-3]s[2]}{s[3a_1+a_2+1]s[3a_1+3a_2]s[3a_1+2a_2]s[3a_1+2a_2+1]s[1]}\right)^{\frac{1}{2}}
\cr
&B[a,u,-2,0,-2,0]=\frac{s[3a_1+2a_2-2-u]s[a_2+u]s[6+u]}{s[3a_1+2a_2-2]s[a_2]s[6]}+\cr
&\frac{s[a_2-3]s[3a_1+2a_2-1]s[3a_1+2a_2+2]s[3a_1+a_2-2-u]s[2]s[u]s[6+u]}{s[a_2]s[3a_1+2a_2]s[3a_1+2a_2-2]s[3a_1+a_2+1]s[1]s[4]s[6]}\cr
&B[a,u,0,-2,-2,0]=\frac{s[u]s[3+u]s[6+u]}{s[1]s[4]s[6]}\times\cr
&\times\left(\frac{s[a_2-2]s[a_2+1]s[3a_1+a_2-1]s[3a_1+a_2+3]s[3a_1+2a_2-1]s[3a_1+2a_2+2]}{s[a_2-1]s[a_2]s[3a_1+a_2+1]^2s[3a_1+2a_2+1]s[3a_1+2a_2]}\right)^{\frac{1}{2}}\cr
&B[a,u,-3,2,-3,2]=\frac{s[6a_1+3a_2-u]s[2+u]s[6+u]}{s[6a_1+3a_2]s[2]s[6]}+\cr
&\frac{s[3a_1+a_2+2]s[3a_1+2a_2+3]s[6a_1+3a_2-2-u]s[u]s[6+u]}{s[3a_1+a_2]s[3a_1+2a_2+1]s[6a_1+3a_2]s[4]s[6]}\cr
&B[a,u,-1,0,-1,0]=\frac{s[3a_1+a_2+u]s[3a_1+2a_2-2-u]s[6+u]}{s[3a_1+a_2]s[3a_1+2a_2-2]s[6]}+\cr
&\frac{s[3a_1+a_2-3]s[3a_1+2a_2-1]s[3a_1+2a_2+2]s[a_2-2-u]s[2]s[u]s[6+u]}{s[3a_1+a_2]s[3a_1+2a_2-2]s[3a_1+2a_2]s[a_2+1]s[1]s[4]s[6]}\cr
&B[a,u,0,-1,-1,0]=\frac{s[u]s[3+u]s[6+u]}{s[1]s[4]s[6]}\cr
&\times\left(\frac{s[a_2-1]s[a_2+3]s[3a_1+a_2-2]s[3a_1+a_2+1]s[3a_1+2a_2-1]s[3a_1+2a_2+2]}{s[a_2+1]^2s[3a_1+a_2-1]s[3a_1+a_2]s[3a_1+2a_2]s[3a_1+2a_2+1]}\right)^{\frac{1}{2}}
}$$
$$\manyeq{
&B[a,u,3,-3,3,-3]=\frac{s[6a_1+4a_2-1+u]s[2+u]s[6+u]}{s[6a_1+4a_2-1]s[2]s[6]}-\cr
&\frac{s[a_2-2]s[3a_1+a_2-2]s[3a_1+2a_2-3]s[6a_1+4a_2+1+u]s[u]s[6+u]}{s[a_2]s[3a_1+a_2]s[3a_1+2a_2-1]s[6a_1+4a_2-1]s[4]s[6]}-\cr
&\frac{G[a+e_3]}{G[a]}\frac{s[6a_1+4a_2+5+u]s[u]s[2+u]}{s[6a_1+4a_2-1]s[4]s[6]}\cr
&B[a,u,2,-2,2,-2]=\frac{s[6a_1+2a_2-1+u]s[2+u]s[6+u]}{s[6a_1+2a_2-1]s[2]s[6]}-\cr
&\frac{s[a_2+2]s[3a_1+a_2-3]s[3a_1+2a_2-2]s[6a_1+2a_2+1+u]s[u]s[6+u]}{s[a_2]s[3a_1+a_2-1]s[3a_1+2a_2]s[6a_1+2a_2-1]s[4]s[6]}-\cr
&\frac{G[a+e_2]}{G[a]}\frac{s[6a_1+2a_2+5+u]s[u]s[2+u]}{s[6a_1+2a_2-1]s[4]s[6]}\cr
&B[a,u,1,-1,1,-1]=\cr
&\frac{s[2a_2-1+u]s[2+u]s[6+u]}{s[2a_2-1]s[2]s[6]}-\cr
&\frac{s[a_2-3]s[3a_1+a_2+2]s[3a_1+2a_2-2]s[2a_2+1+u]s[u]s[6+u]}{s[a_2-1]s[3a_1+a_2]s[3a_1+2a_2]s[2a_2-1]s[4]s[6]}-\cr
&\frac{G[a+e_1]}{G[a]}\frac{s[2a_2+5+u]s[u]s[2+u]}{s[2a_2-1]s[4]s[6]}\cr
&B[a,u,0,0,0,0]= \frac{s[6-u]s[12+u]s[3+u]}{s[6]s[12]s[3]}+\frac{s[u]s[3+u]s[6+u]}{s[6]s[9]s[12]}\times\cr
&\times\sum _{\mu =1}^3 \left(\frac{G[a+e_\mu]}{G[a]}B[a,-12,0,\mu,0,\mu ]+\frac{G[a+e_{-\mu}]}{G[a]}B[a,-12,0,-\mu ,0,-\mu ]\right)
}$$
Taking symmetries (eq.(9a) in ref. \r\Kuniba) into account, we get also
$$\manyeq{
&B[a,u,-3,-2,-2,-3]=B[a,u,-2,-3,-3,-2]\cr
&B[a,u,-3,-1,-1,-3]=B[a,u,-1,-3,-3,-1]\cr
&B[a,u,-2,1,1,-2]=B[a,u,1,-2,-2,1]\cr
&B[a,u,-3,0,0,-3]=B[a,u,0,-3,-3,0]\cr
&B[a,u,0,-3,-2,-1]=B[a,u,-2,-1,0,-3]\cr
&B[a,u,0,-3,-1,-2]=B[a,u,-1,-2,0,-3]\cr
&B[a,u,0,-2,1,-3]=B[a,u,1,-3,0,-2]\cr
&B[a,u,-2,0,0,-2]=B[a,u,0,-2,-2,0]\cr
&B[a,u,-1,0,0,-1]=B[a,u,0,-1,-1,0]
}$$

Taking symmetry  eq.(9b) of ref. \r\Kuniba\ into account
$$\manyeq{
 B[a,u,\kappa,\eta,\mu,\nu]=B[-a,u,-\kappa ,-\eta ,-\mu ,-\nu ]
}$$
Notice the difference with respect to  eq.(9.b) of ref. \r\Kuniba, where an additional factor $2\rho$ is present in the RHS, $-a\rightarrow -a-2\rho$.

And also the symmetry eq.(9c)  
$$\manyeq{
B[a,u,\kappa,\eta,\mu,\nu]=-(-1)^{\kappa -\nu }\sqrt{\frac{G[a+e_\kappa]G[a+e_\mu ]}{G[a]G[a+e_\kappa
+e_\eta]}}B[a+e_\mu,-6-u,-\mu ,\kappa ,\nu ,-\eta ]
 }$$

\endpage

\bye

\bye